\documentclass[10pt, journal]{IEEEtran}
\usepackage{blindtext}
\usepackage{amsmath}
\usepackage{amssymb}
\usepackage{bm}
\usepackage{mdwmath}
\usepackage{lipsum}
\usepackage{graphicx}
\usepackage{xcolor}
\usepackage{subfigure}
\usepackage{stfloats}
\usepackage{makecell}
\usepackage{multirow}
\usepackage{acronym}
\usepackage{enumerate}
\usepackage[noadjust]{cite}
\usepackage{url}
\usepackage{algorithm}
\usepackage{algpseudocode}
\newcounter{MYalgorithmic}
\renewcommand{\theMYalgorithmic}{\arabic{MYalgorithmic}}
\newcommand{\algcaption}[1]{
	\refstepcounter{MYalgorithmic}
	\textbf{Algorithm}~\textbf{\theMYalgorithmic}.~#1}
\newenvironment{MYalgorithmic}[5]
{
	\hrule height 1.2pt
	\vspace{3pt}
	#1{#2}%
	#3{#4}
	\vspace{3pt}
	\hrule height 0.5pt
	\vspace{3pt}
	#5
}
{
	\vspace{3pt}
	\hrule height 0.5pt
}
\newtheorem{definition}{\textbf{\textit{Definition}}}

\newtheorem{theorem}{\textbf{\textit{Theorem}}}
\newtheorem{remark}{\textbf{\textit{Remark}}}

\newtheorem{challenge}{\textbf{\textit{Challenge}}}

\newcommand{\upperroman}[1]{\uppercase\expandafter{\romannumeral#1}}

\usepackage{cases}
\usepackage{bbm}
\usepackage{booktabs}
\begin{document}
\title{Semi-Decentralized Network Slicing for Reliable V2V Service Provisioning: A Model-free Deep Reinforcement Learning Approach}

\author{Jie Mei,~\IEEEmembership{Member,~IEEE,}
		Xianbin Wang$^{*}$,~\IEEEmembership{Fellow,~IEEE},
		Kan Zheng,~\IEEEmembership{Senior Member,~IEEE}
\thanks{Manuscript received January 20, 2021; revised March 11, 2021, accepted July 19, 2021. This work was supported in part by the Natural Sciences and Engineering Research Council of Canada (NSERC) Discovery Program under Grant RGPIN2018-06254 and in part by the Canada Research Chair Program. (\textit{Corresponding author: Dr. Xianbin Wang}.)}
\thanks{Jie~Mei and Xianbin~Wang are with the Department of Electrical and Computer Engineering, Western University, London, ON N6A 5B9, Canada, (E-mails: \{jmei28, xianbin.wang\}@uwo.ca).}
\thanks{Kan~Zheng is with the the Intelligent Computing and Communication ($ \text{IC}^\text{2} $) Lab, Beijing University of Posts and Telecommunications (BUPT), Beijing, 100876, China, (E-mail: zkan@bupt.edu.cn).}
}
\maketitle

\begin{abstract}
  Applying of network slicing in vehicular networks becomes a promising paradigm to support emerging Vehicle-to-Vehicle (V2V) applications with diverse quality of service (QoS) requirements. However, achieving effective network slicing in dynamic vehicular communications still faces many challenges, particularly time-varying traffic of Vehicle-to-Vehicle (V2V) services and the fast-changing network topology. By leveraging the widely deployed LTE infrastructures, we propose a semi-decentralized network slicing framework in this paper based on the C-V2X Mode-4 standard to provide customized network slices for diverse V2V services. With only the long-term and partial information of vehicular networks, eNodeB (eNB) can infer the underlying network situation and then intelligently adjust the configuration for each slice to ensure the long-term QoS performance. Under the coordination of eNB, each vehicle can autonomously select radio resources for its V2V transmission in a decentralized manner. Specifically, the slicing control at the eNB is realized by a model-free deep reinforcement learning (DRL) algorithm, which is a convergence of Long Short Term Memory (LSTM) and actor-critic DRL. Compared to the existing DRL algorithms, the proposed DRL neither requires any prior knowledge nor assumes any statistical model of vehicular networks. Furthermore, simulation results show the effectiveness of our proposed intelligent network slicing scheme.
\end{abstract}


\section{Introduction}
\label{sec:Introduction}
In recent years, vehicle-to-vehicle (V2V) communication has become one critical enabler for the rapidly growing connected vehicle and intelligent transportation industries. The global connected vehicle market is expected to grow from \$ 42.25 billion in 2018 to \$ 142.49 billion by 2026, expanding at a Compound Annual Growth Rate (CAGR) of 16.4 \% \cite{Kenneth}. Meanwhile, emerging V2V applications, such as cooperative collision avoidance, autonomous driving, and platooning control, have led to a broad spectrum of Quality of Service (QoS) requirements on vehicular networks \cite{R2, R3}. However, conventional vehicular networks supporting human-centric applications cannot fully meet the highly diverse QoS requirements of future V2V applications. This limitation requires the new generation of vehicular networks to enable diverse QoS provisioning by intelligent and efficient utilization of limited radio resources \cite{R4}.
\par
One promising solution for diverse QoS provisioning in vehicular networks is network slicing, which provides a multipurpose platform to enable a wide range of applications and services. Network slicing creates multiple virtual customized networks, referred to as network slices, on top of a common substrate infrastructure. Therefore, the main goal of this paper is to implement the network slicing paradigm in vehicular networks, where the operator can flexibly compose network slices for meeting specific QsS demands of various V2V applications. 
\par
However, the detailed design of the network slicing scheme for vehicular networks is still very challenging due to the following two open issues,
\begin{itemize}
    \item How to integrate the network slicing paradigm with state-of-art vehicular communication techniques in a cost-effective and scalable manner?
    \item How to realize proactive and situation-aware network slicing that can ensure diverse QoS requirements of V2V services in time-varying vehicular networks?
\end{itemize}
This paper addresses these two issues by combining the Cellular Vehicle-to-Everything (C-V2X) system with the concept of Artificial Intelligence (AI) empowered network slicing.
\par
Firstly, the C-V2X standard has been proposed to replace the existing IEEE 802.11p protocol, which cannot fully support today’s V2V services. Currently, the C-V2X standard includes two modes of operation, i.e., C-V2X Mode-3 and C-V2X Mode-4. In Mode-3, eNodeB (eNB) directly allocates radio resources to vehicles for their V2V transmissions in a centralized way. In Mode 4, vehicles perform distributed radio resource scheduling to autonomously select radio resources from a radio resource pool without a centralized scheduler \cite{R7}. However, compared with Mode 4, Mode 3 could cause unbearable control signaling overhead and processing delay in the dense and dynamic V2V scenarios. Since the poor scalability of Mode 3, exploration of Mode 4 is a potential direction to enable diverse QoS provisioning in realistic transportation environments.
\par
Secondly, although network slicing can customize network slices according to specific QoS demands, this performance gain comes at the cost of introducing much more complexity into the communication system. This high complexity makes traditional mathematical model-based approaches to awareness of network situation and network operation no longer adequate since the model-based approaches either lack explicit models or do not have the processing time to calculate heuristic solutions \cite{R1,TCOM4}. This challenge motivates us to propose an AI-empowered network slicing architecture for vehicular networks to support vehicular applications \cite{Mei}. It shows great potential in developing intelligent network slicing schemes for supporting V2V applications with diverse QoS requirements.
\par
With the observed considerations, this paper proposes a semi-decentralized network slicing framework based on C-V2X Mode 4 to maximize the long-term QoS performances of V2V services. In principle, the operation of network slices is under the supervision of eNB. Based on deep reinforcement learning (DRL), eNB extracts the underlying network situations and adjusts slice configuration accordingly. Under the coordination of eNB, vehicles of each slice are automatically performing radio resource scheduling procedures. The following briefly summarizes the main technical contributions of this work:
\begin{itemize}
    \item Design of a semi-decentralized network slicing framework based on C-V2X Mode 4. It has two layers. First, eNB executes adaption of slice configuration, i.e., inter-slice radio resource allocation and tuning of parameters of Mode 4 protocol, according to the network dynamics at a large timescale. Then, conditioned by the slice configuration determined by the eNB, the vehicle in each slice performs autonomous radio resource selection for V2V transmission based on Mode 4. In this framework, the eNB performs slicing control with coarse resource and time granularity and does not require frequent interaction between eNB and vehicles, significantly reducing the signaling overhead and allowing sufficient time for intelligent processing.
    \item Model-free DRL to realize the situation-aware slicing control with only partial information of vehicular networks. The adaption of slice configuration at eNB is functioned by a DRL agent. However, the eNB only has partial observation information of vehicular networks. This makes conventional DRL methods inefficient, which are relying on the prior knowledge of systems. Therefore, based on the partially observed Markov decision process (PoMDP), we propose an actor-critic structured DRL algorithm by exploring the long short-term memory (LSTM). Specifically, LSTM enables the eNB to extract the underlying network situation from historical, partial information of the vehicular network. With the proposed DRL algorithm, eNB can perform slicing control with self-configuration and self-optimization capabilities.
\end{itemize}
\par
The remainder of this paper is structured as follows. Section II presents related works. Section III describe the considered system model of network slicing in Mode 4 based vehicular networks. In Section IV, we propose a semi-decentralized network slicing framework based on C-V2X Mode 4. In Section III, we formulate the optimization of slicing configuration policy at the eNB as a PoMDP problem. In Section IV, we propose an actor-critic structured DRL algorithm to solve the formulated problem. In Section V, we present numerical experiments to compare the performance of the proposed algorithm against state-of-the-art baseline schemes. Finally, Section VI draws the conclusions.
\section{Related Works}
Since the C-V2X standard is relatively new, which is introduced in 3GPP Releases 14 and 15 and will be further enhanced in Release 16 \cite{3gpp01,3gpp02}. Current works mainly focus on three research topics: performance analysis of C-V2X network, radio resource management, and network slicing in C-V2X networks.
\par
As mentioned above, C-V2X Mode-4 employs the distributed radio scheduling scheme, referred to as sensing based Semi-Persistent Scheduling (SPS) scheme, to enable autonomous radio resource management of each vehicle. In the sense-based SPS scheme, vehicles sense and keep a history of the channel status and utilize it to select suitable radio resources for V2V transmissions. Since the autonomy nature of C-V2X Mode 4, it faces radio resource sharing conflicts, i.e., packet collisions when two or more vehicles simultaneously utilize the same radio resources. This issue will affect the performance of vehicular networks. Thus, based on probabilistic theory, performance analytical models of C-V2X Mode 4 network are proposed for quantifying the collision probability, and throughput as a function of vehicle density and the distance between transmitting and receiving vehicle \cite{R12,R13}. Besides, based on network-level simulations, the authors in \cite{R141,r15,R16} analyze the impact of the main parameters of Mode 4 on the network performance, which shows that Mode 4 is robust and scalable for highly dynamic vehicular scenarios.
\par
To further improve the performance of C-V2X networks, different studies have proposed options to enhance the radio resource management schemes. The authors in \cite{R17} propose a distributed radio resource management scheme for C-V2X Mode 4, which exploits geography information of vehicles to improve V2V communication reliability. Likewise, in \cite{R18}, a spatial reuse-based radio resource management scheme is proposed to improve the spectrum utilization of vehicular networks. Instead of simply improving spectrum utilization, in the context of C-V2X assisted autonomous driving, the authors in \cite{R20} jointly optimize radio resource allocation, cooperative driving perception, and vehicle controls to improve driving safety and transportation efficiency.
\par
However, the effectiveness of C-V2X Mode 4 still needs to be improved due to the following two limitations:
\begin{itemize}
    \item Low operation efficiency of C-V2X Mode 4 network. In this decentralized network, each vehicle performs the distributed radio resource scheduling independently based on its local knowledge and needs, leading to selfish decisions from different vehicles. Thus, to avoid unreasonable decisions from each vehicle, it is critical to establish coordination between eNB and vehicles.
    \item Extreme difficult for network situational awareness. Firstly, due to the high mobility of vehicles, vehicular network status could change rapidly. Meanwhile, real-time sensing of network situations will consume excessive overheads to exchange sensing information between eNB and vehicles. These make the real-time and precise awareness of network situations challenging at the eNB.
\end{itemize}
Recently, network slicing has been introduced into vehicular networks to meet diverse QoS requirements for V2X services. By leveraging Lyapunov optimization, \cite{TVT_3} proposes a RAN slicing scheduling strategy for the joint radio resource allocation and power control, aiming to maximize long-term network capacity while guaranteeing the strict QoS requirements of V2V services. In \cite{TWC_3}, a hierarchical RAN slicing framework is developed for the heterogeneous vehicular networks, where other slices opportunistically reuse the idle radio resources of one network slice to improve the spectrum efficiency.
\par
Based on our literature review and analysis, the above works on the C-V2X based vehicular networks have the following two limitations,
\begin{itemize}
    \item Most of the literature assumes that network infrastructures can fully observe the status of the vehicular network. However, this assumption is too optimistic for the real scenarios of vehicular networks due to the high mobility of vehicles \cite{R24}. This characteristic hinders the direct sensing of network situations since it needs frequent interaction between eNB and vehicles, which will consume a large amount of signaling overheads. Thus, one effective solution is to enable the network slicing with only the long-term and partial information of vehicular networks.
    \item Most works are regulated by the conventional mathematical model-based approaches. The fundamental premise of these model-based approaches is to obtain a precise mathematical model to describe the system. Then, based on the accurate system model, we can further analyze or optimize the system performance. However, the dynamics and variation pattern of vehicular networks are difficult to be modeled accurately. Thus, it is reasonable for us to enable the operation of C-V2X based networks through a model-free AI technology, such as model-free DRL technologies. 
\end{itemize}
\textit{Notations}: In the following, italic boldface lower-case and upper-case characters denote vectors and matrices, respectively. Sets are denoted by calligraphic letters, i.e., $\mathcal{U}$. The operator $\left| {\cal U} \right|$ represents the cardinality of set $\mathcal{U}$. To ease readability, we list the major notations in Table I.
\begin{figure*}[!t]
	\centering
	\setlength{\abovecaptionskip}{0.cm}
	\includegraphics[width=0.65\textwidth]{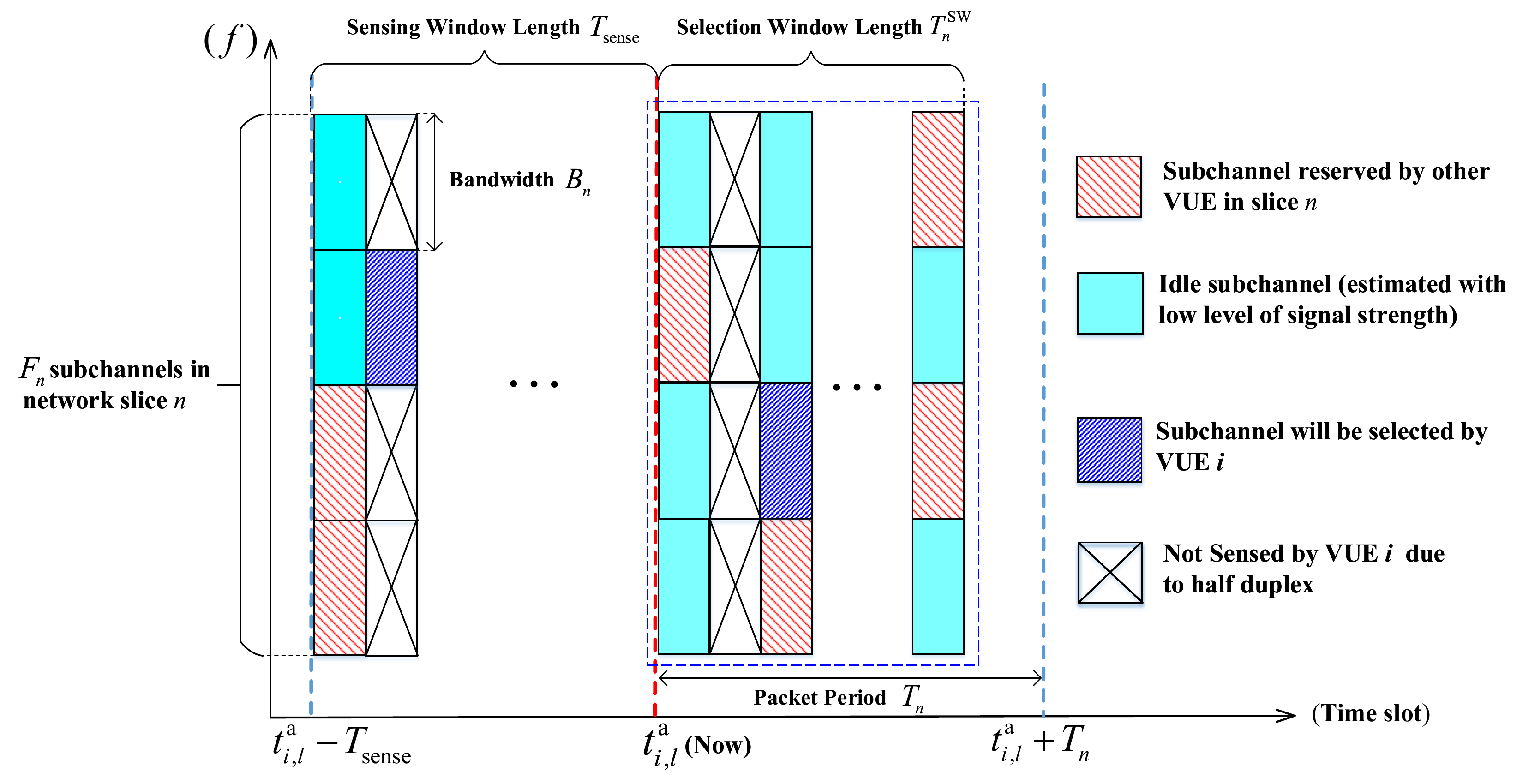}  
	\caption{ {An illustration of Sensing based Semi-persistent scheduling (SPS) process of VUE \begin{small}$i \in \mathcal{V}_n$\end{small} in network slice $n$: At slot $t_{i,l}^{\rm{a}}$, VUE \begin{small}$i$\end{small} needs to select a sub-channel to transmit its packet. VUE \begin{small}$i$\end{small} can select a subchannel within a Selection Window (SW). SW is a time window that ranging from slot \begin{small}$t_{i,l}^{\rm{a}}$\end{small} to \begin{small}$t_{i,l}^{\rm{a}}+T_n^{\rm{sw}}$\end{small}, where \begin{small}$T_n^{\rm{sw}}$\end{small} is the length of SW. Then, VUE \begin{small}$i$\end{small} sorts the idle sub-channels in terms of the average signal strength, which are sensed during the last sensing window that ranging from slot \begin{small}$t_{i,l}^{\rm{a}}-T_{\rm{sense}}$\end{small} to slot \begin{small}$t_{i,l}^{\rm{a}}$\end{small}. Finally, VUE \begin{small}$i$\end{small} selects a sub-channel with the lowest average signal strength.}}
	\label{Fig1}
	\vspace{-0.3cm}
\end{figure*}
\begin{table}
    \centering
	\footnotesize
	\renewcommand{\arraystretch}{1.1}
	\caption{ {Major Notations used in this Paper}}
	\label{Notation_Table}
	\begin{tabular}{cl}
	    \toprule
		\textbf{Notation} & \textbf{Definition} \\
		\hline
		${\mathcal N}$ & Set of network slices: $\left\{ {1,...,N} \right\}$ \\
		${\mathcal V}_n$ & Set of VUEs belong to network slice $n \in \mathcal{N}$ \\
		$B$ & Total available bandwidth \\
		$\delta$ & Duration of each time slot \\
		$F_n$ & Number of subchannel in slice $n \in \mathcal{N}$ \\
		$B_n$ & Bandwidth of each subchannel in slice $n \in \mathcal{N}$ \\
		$T_n^{\rm{sw}}$ & Length of the selection window in slice $n \in \mathcal{N}$ \\
		$s_{i,m,t}$ & $s_{i,m,t}=1$ if VUE $i$ selects the $m$-th sub-channel at slot $t$ \\
		$r_{i,t}$ & Data rate of VUE $i$ at slot $t$ \\
		$\Delta t$ & Each epoch is composed by $\Delta t$ consecutive slots\\
		${\bar d}_{n,k}$ & Average packet delay of VUEs in slice $n$ \\
		${\bar \beta}_{n,k}$ & Average Packet Drop Ratio (PDR) of slice $n$ \\
		${\bar x}_{n,k}$ & Average subchannel occupancy ratio in slice $n$ \\
		${\boldsymbol{O}_{n,k}}$ & Observation of slice $n$ at epoch $k$: $\left\{ {{\left| {{{\mathcal V}_{n,k}}} \right|},{\bar d}_{n,k},{\bar \beta}_{n,k}, {\bar x}_{n,k}} \right\}$\\
		${\boldsymbol{O}_{k}}$ & Observation of network at epoch $k$: $\{ {{{\boldsymbol{O}}_{n,k}}\left| {n \in {\mathcal N}} \right.}\}$\\
		${\boldsymbol{H}_{k}}$ & Observation history of network at epoch $k$: $\left( {{{\boldsymbol{O}}_1},...,{{\boldsymbol{O}}_{k - 1}}} \right)$\\
		${\boldsymbol{C}_k}$ & Configuration of slice $n$ at epoch $k$: $\left\{ {{F_n},{B_n},T_n^{{\rm{sw}}}, \forall n \in {\cal N}} \right\}$\\
		${\pi}_{\boldsymbol \theta}$ & Slice Configuration Policy: Mapping from ${\boldsymbol{H}_{k}}$ to ${\boldsymbol{C}_k}$\\
		$J\left(\boldsymbol{O}_k\right)$ & QoS-related reward function of all slices\\
		\bottomrule
	\end{tabular}
	\vspace{-0.3cm}
\end{table}
\section{System Model}
Consider a freeway scenario with one eNB,  {where total available bandwidth is $B$}.  {The time dimension is partitioned into slots of duration $\delta$, indexed by $t \in \left\{1,2,...\right\}$. Assume the physical vehicular network is split into $N$ network slices, denoted by $\mathcal{N}=\{1,2,...,N\}$, each of which has a specific V2V application it provides.} Vehicular user equipment (VUE), counted in terms of transmitter, associated with slice $n\in\mathcal{N}$ are denoted as $\mathcal{V}_n$. Assumed that all VUEs can successfully receive the information of slice configuration from the eNB. 
\subsection{Traffic Model of V2V Services}
 In slice $n\in\mathcal{N}$, each VUE needs to periodically transmit packet to its receiving vehicle with a period of $T_n$ slots. Assume the packet in slice $n \in \mathcal{N}$ have a fixed data size $Z_n$ (in bits). It is assumed that each VUE has a queue buffer to store the packet to be delivered, and the packet is delivered based on the first-come-first-serve (FCFS) criteria.  Let $l$($= 1,2,...$) denotes the index of the packet arriving at VUE $i$’s buffer in slice $n$. Furthermore, the inter-packet arrival time instant of $l$-th packet of VUE $i$ is denoted as $t_{i,l}^{\rm{a}} = l \cdot T_n$.
\subsection{V2V communication based on C-V2X Mode 4}
The total bandwidth $B$ are sliced and assigned to each slice by the eNB. In slice $n\in\mathcal{N}$, the assigned bandwidth are re-organized as $F_n$ sub-channels, indexed by $m \in \left\{1,2,...,F_n\right\}$, and the bandwidth of each subchannel is $B_n$ ($F_n B_n < B$).
\par
In Mode 4, VUE autonomously selects and reserves sub-channel for V2V transmission with Sensing based Semi-Persistent Scheduling (SPS) scheme~\cite{R7}. As shown in Figure \ref{Fig1}, at slot $t_{i,l}^{\rm{a}}$, supposing that VUE \begin{small}$i \in \mathcal{V}_n$\end{small} needs to select a sub-channel to transmit $l$-th packet ($l= 1,2,...$), then this procedure can be divided into the following three steps,
\textbf{Step 1 (Sensing)}: VUE $i$ continuously senses the signal strength in each subchannel during the last \begin{small}$T_{\rm{sense}}$\end{small} slots before slot $t_{i,l}^{\rm{a}}$ (referred to as sensing window), and calculate the average signal strength of subchannel.

\textbf{Step 2 (Subchannel Selection and Transmission)}:  {VUE $i$ can select a subchannel within a Selection Window (SW). SW is a time window that includes the slots in the range \begin{small}$\left[t_{i,l}^{\rm{a}},t_{i,l}^{\rm{a}}+T_n^{\rm{sw}}\right]$\end{small}, where \begin{small}$T_n^{\rm{sw}}$\end{small} is the length of selection window}. VUE $i$ sorts the candidate sub-channels in terms of the average received signal strength during the last sensing window, and then reserves the sub-channel with the lowest average received signal strength, which can be stated as:
\begin{itemize}
    \item VUE $i$ ranks all sub-channels in the selection window by their average signal strength in a descending order and selects the bottom 20\% of them to compose the list of candidate sub-channels, denoted as $\mathcal{S}$;
    \item VUE $i$ will randomly choose one of the candidate sub-channel in list $\mathcal{S}$. Assume VUE $i$ chooses $m$-th (\begin{small}$1 \le m \le F_{n}$\end{small}) sub-channel at $t$-th slot for transmitting $l$-th packet. Let $s_{i,m,t}$ denote sub-channel selection indicator for VUE $i$, where $s_{i,m,t}=1$ means $m$-th sub-channel at the $t$-th slot is chosen by VUE $i$; otherwise, $s_{i,m,t}=0$.
\end{itemize}
However, $m$-th sub-channel at $t$-th slot may be used by other VUEs in slice $n$. For instance, supposing that VUE $i$ and \begin{small}$i'\in \mathcal{V}_n$\end{small} simultaneously selects $m$-th sub-channel at $t$-th slot, i.e., $s_{i,m,t}=s_{i',m,t}=1$. This condition will induce the intra-slice interference, which deteriorates the reliability of V2V communication. Then, Signal-to-Interference-plus-Noise Ratio (SINR) of the receiving vehicle of VUE $i$ in $m$-th subchannel at $t$-th slot is given by\begin{small}
    \begin{equation}
    \gamma _{i,m,t} = \frac{{{P}{{\left| {g_{i,m,t}} \right|}^2}}}{{\sum\nolimits_{i' \in {{\cal V}_n}{\backslash}\left\{i\right\}} {{s_{i',m,t}}{P}{{\left| {g_{i',m,t}} \right|}^2}}  + {N_0}}},\;{\rm if}\;s_{i,m,t}=1,
    \label{Eq_1}
\end{equation}
\end{small}where \begin{small}$g_{i,m,t}$\end{small} is the channel gain, which contains path loss, shadowing effect and small-scale fading, from VUE $i$ to its receiving vehicle in $m$-th subchannel, \begin{small}$g_{i',m,t}$\end{small} is the interference channel gain from VUE $i'$ to the receiving vehicle of VUE $i$, $P$ is the transmitted power of VUE, and $N_0$ is the power of additive white Gaussian noise (AWGN) in each sub-channel. \par
Therefore, the achievable data rate of receiver of VUE $i$ at the $t$-th slot can be approximated by Shannon theory,
\begin{small}
\begin{equation}
    {r_{i,t}} = \sum\nolimits_{m = 1}^{{F_n}} {{s_{i,m,t}}\cdot \left[ { {B_n\delta\cdot \log } \left( {1 + \gamma _{i,m,t}} \right)} \right]} ,\;i \in {{\cal V}_n},
\end{equation}
\end{small}where $B_n$ is the bandwidth of each subchannel in slice $n$ and $\delta$ is the time duration of slot.
\par
The selected sub-channel is used to transmit a full packet. Then, delay of $l$-th packet at VUE $i$’s buffer can be expressed as,\begin{small}
\begin{equation}
    {d_{i,l}}{\rm{\;=\;}}t - t_{i,l}^{\rm{a}},\;i \in {{\cal V}_n}.
    \label{eq_3}
\end{equation}
\end{small}Specifically, packet latency ${d_{i,l}}$ approximately follows discrete uniform distribution \begin{small}
${\rm unif}\left\{1,T_n^{\rm{sw}}\right\}$\end{small}. Thus, in slice $n \in \mathcal{N}$, the latency of VUE is impacted by \begin{small}$T_n^{\rm{sw}}$\end{small} (selection window length).
\par
In our system model, packet is lost when $Z_n$ error-free bits (i.e., packet size) cannot be correctly decoded by the receiving vehicle of VUE $i$. Thus, let binary variable $L_{i,l}$ denotes the packet loss indicator at VUE $i$’s buffer, which can be represented as\begin{small}
\begin{numcases}
{L_{i,l} = }
{1,\;{\rm if}\;{r_{i,t}} < {Z_n}}, \notag\\
{0,\;{\rm otherwise} \notag}
\end{numcases}
\end{small}\par
Furthermore, we define subchannel occupancy ratio, $x_{n,t}$, to characterize the level of subchannel congestion in slice $n$, \begin{small}
\begin{equation} \label{x_t}
    x_{n,t} = {{\sum\nolimits_{m = 1}^{{F_n}} {\mathbbm{1}\left\{ {\sum\nolimits_{i \in {{\cal V}_n}} {{s_{i,m,t}}}  \ge 1} \right\}} } \mathord{\left/
 {\vphantom {{\sum\nolimits_{m = 1}^{{F_n}} {\mathbbm{1}\left\{ {\sum\nolimits_{i \in {{\cal V}_n}} {{s_{i,m,t}}}  \ge 1} \right\}} } {{F_n}}}} \right.
 \kern-\nulldelimiterspace} {{F_n}}},
\end{equation}
\end{small}where $\mathbbm{1}\left\{\cdot\right\}$ is the indicator function. 
\textbf{Step 3 (Reservation and Re-selection)}: Once a sub-channel is reserved, the same sub-channel will be used for several consecutive V2V transmissions. After a random number of V2V transmissions VUE $i$ will reselect its reserved sub-channel with probability $p_{\rm{res}}$, and repeat \textbf{Step 1} and \textbf{2}.
\begin{figure*}[!t]
\centering
\includegraphics[width=0.8\textwidth]{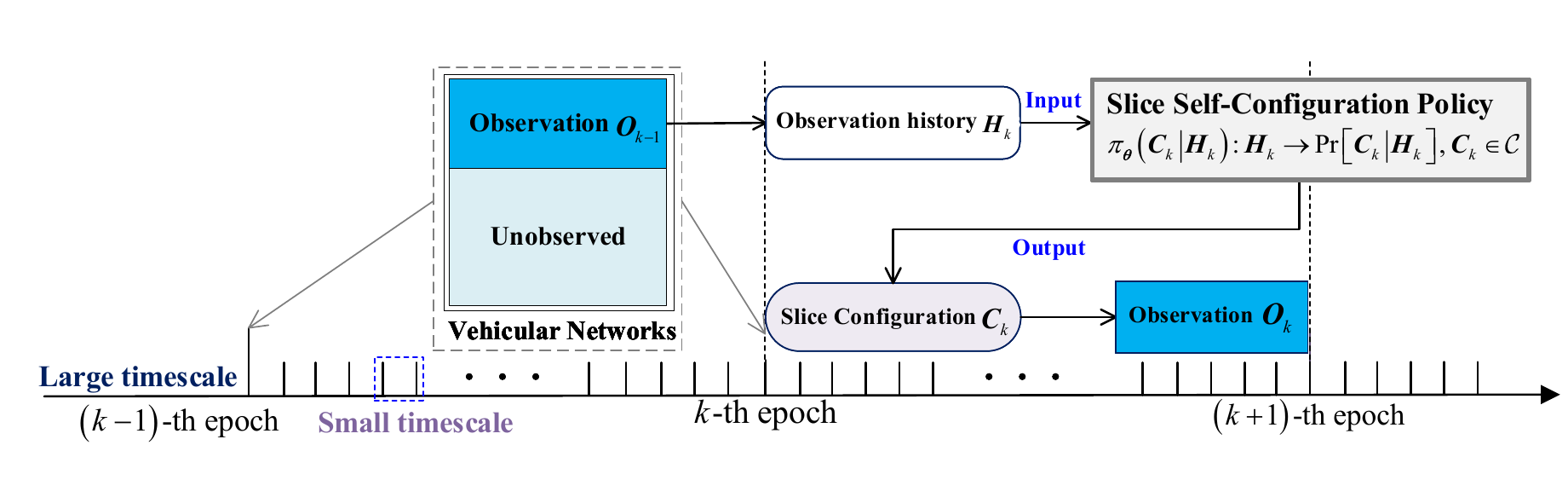}
\caption{ {The flowchart of the proposed semi-decentralized network slicing scheme for the C-V2X Mode 4 based vehicular networks.}}
\label{Fig2}
\end{figure*}
\begin{remark}
Specifically, all VUEs associated with network slice $n \in \mathcal{N}$ have the same parameters (i.e., $F_n$, $B_n$ and \begin{small}$T_n^{\rm{sw}}$\end{small}) of C-V2X Mode 4. The V2V communication performance is determined by these parameters. In our proposed network slicing framework (Section III-A), the eNB will determine how to adjust the parameters of C-V2X mode 4 for each slice according to the network situations.
\end{remark}
\section{Semi-decentralized Network Slicing and Problem Formulation}
\subsection{Proposed Network Slicing Framework}\label{3-1}

In this study, as an extension of our original work~\cite{Mei}, a semi-decentralized network slicing framework for vehicular networks is proposed. It consists of an upper-level and a lower-level. As shown in Figure 2, at the upper-level, eNB is responsible for adjusting the slice configuration according to the dynamics of V2V service traffic at a large timescale. Specifically, the time resolution of upper-level is defined as an epoch, indexed by $k\in\{1,2,...\}$, the $k$-th epoch is ranging from slot \begin{small}$k \cdot \Delta t$\end{small} to slot \begin{small}$(k+1)\cdot{\Delta t}-1$\end{small}. At the lower-level, vehicles in each slice autonomously select their sub-channel using the sensing-based SPS scheme, which is configured by the upper-level. It is noteworthy that the upper-level is not involved in the real-time radio resource scheduling for V2V communications.
\par
The upper-level is in charge of tuning the slice configuration to improve the QoS performance of services according to the partial observation history of vehicular networks. First, the specific definition of partial observation of vehicular networks is given as follow:
\begin{definition}[\textbf{Partial Observation of Vehicular Networks}]
\label{def1}
The observation of vehicular network at $k$-th epoch is defined as
\begin{small}
\[{{\boldsymbol{O}}_k}{\rm{ = }}\left\{ {{{\boldsymbol{O}}_{n,k}}\left| {n \in {\mathcal N}} \right.} \right\} \in {\mathcal O},\]
\end{small}where the observation of slice $n \in \mathcal{N}$ at $k$-th epoch is
\begin{small}
\[{{\boldsymbol{O}}_{n,k}}{\rm{ = }}\left\{ {{\left| {{{\mathcal V}_{n,k}}} \right|}, {\bar x}_{n,k}} \right\},\]
\end{small}
\begin{enumerate}[a).]
\item \textit{Number of VUEs in slice} $n$,\begin{small}
$\left| {{{\mathcal V}_{n,k}}} \right|$\end{small}: It is defined as the number of vehicles associated with slice $n$ in $k$-th epoch, i.e. \begin{small}$\left| {{{\mathcal V}_{n,k}}} \right|$\end{small}, where \begin{small}$\mathcal{V}_{n,k}$\end{small} is the VUE set in slice $n$ during $k$-th epoch. Because epoch is in the level of hundreds of milliseconds, which is smaller than the vehicle inter-arrival time. Therefore, we assume that set \begin{small}${{{\mathcal V}_{n,k}}}$\end{small} does not change within $k$-th epoch.
\item Average subchannel occupancy ratio in slice $n$, ${\bar x}_{n,k}$: It is defined as\begin{small}
\begin{equation}
    {{\bar x}_{n,k}} = \frac{1}{{\Delta t}}\sum\nolimits_{t = k\Delta t}^{\left( {k + 1} \right)\Delta t - 1} {{x_{n,t}}}.\notag
\end{equation}
\end{small}
\end{enumerate}
Besides, \begin{small}${\mathcal O}$\end{small} is the set of all possible observations of vehicular networks.
\end{definition}
\begin{remark}
Within each epoch, the eNB gets these two kinds of information through sensing subchannels at each slot, which barely needs information exchange between eNB and vehicles in its service area. At the end of each epoch, eNB aggregates the temporal dynamics of these raw information within epoch and converted into the observation of vehicular network \begin{small}${\boldsymbol O}_{k}$\end{small}. However, due to the high complexity and time varying nature of vehicular networks as well as the limited sensing ability, the observation of vehicular networks \begin{small}${\boldsymbol O}_k$\end{small} is an \textit{partial information} of network situation, which cannot be regarded as a full status of vehicular network.
\end{remark}
\par
Therefore, the eNB (i.e., the upper-level controller) can exploit the observation history of vehicular networks and infer the full information of network situation. Specifically, 
\begin{definition}[\textbf{Observation History of Vehicular Networks}]
\label{def3}
At the beginning of $k$-th epoch, the eNB obtains the partial observation of vehicular networks during the previous epoch (i.e., \begin{small}${\boldsymbol O}_{k-1}$\end{small}) and add it to the observation history. Herein, the observation history of vehicular networks is defined as
\begin{small}
\[{{\boldsymbol H}_k} = \left( {{{\boldsymbol{O}}_1},...,{{\boldsymbol{O}}_{k - 1}}} \right).\]
\end{small}\end{definition}\par
With the observation history, as shown in Figure~\ref{Fig2}, the upper level controlling policy are described as follows,
\begin{definition}[\textbf{Network Slicing Configuration Policy}]
\label{def4}
The slicing configuration policy ${\pi}$ is defined as a stochastic policy,\begin{small}
\[{\pi _\theta }\left( {{{\boldsymbol C}_k}\left| {{{\boldsymbol H}_k}} \right.} \right):{{\boldsymbol H}_k} \to \Pr \left[ {{{\boldsymbol C}_k}\left| {{{\boldsymbol H}_k}} \right.} \right],\;{{\boldsymbol C}_k} \in {\mathcal{C}},\]
\end{small}which is a mapping from observation history of vehicular networks \begin{small}${\boldsymbol H}_k$\end{small} to a probability distribution over the candidate slice configuration $\mathcal{C}$, which is the collection of all candidate slice configurations. Furthermore, \begin{small}${\boldsymbol C}_k$\end{small} is the slice configuration at $k$-th epoch, \begin{small}
\begin{equation}
    {{\boldsymbol C}_k}{\rm{ = }}\left\{ {{F_n},{B_n},T_n^{{\rm{sw}}}\left| {\sum\nolimits_{n \in {\cal N}} {{{\cal F}_n} \cdot {B_n}}  \le {B},n \in {\cal N}} \right.} \right\}, \notag
\end{equation}
\end{small}
\begin{enumerate}[a).]
    \item $F_n$ is the number of subchannels in slice $n$;
    \item $B_n$ is the number of subchannels in slice $n$; 
    \item \begin{small}$T_n^{\rm{sw}}$\end{small} is the length of selection window of slice $n$. 
\end{enumerate} 
\end{definition}
\begin{remark}
In this paper, the network slicing configuration policy is represented by an Artificial Neural Network (ANN) ${\pi}_{\boldsymbol{\theta}}$, where $\boldsymbol{\theta}$ is the weight vector associated with this ANN. The output of ${\pi}_{\boldsymbol{\theta}}$ is a probability distribution over candidate slice configurations. For instance, \begin{small}${\pi _\theta }\left( {{{\boldsymbol C}_k}\left| {{{\boldsymbol H}_k}} \right.} \right)$\end{small} is the probability of selecting slice configuration \begin{small}${\boldsymbol{C}}_k \in {\mathcal{C}}$\end{small} under the condition of observation history \begin{small}${\boldsymbol{H}}_k$\end{small}.
\end{remark}

\subsection{Problem Formulation}
 {In the following, we formulate the optimization of the RAN slicing policy in the proposed scheme as a stochastic optimization problem, whose goal is to maximize the long-term QoS performance of V2V network slices. Specifically, the average packet delay of UE, should be considered as one important metric of QoS for each slice, which is defined as,}\begin{small}
\begin{equation}
    {\bar d_{n,k}}{\rm{ = }}\frac{1}{{{\left| {{{\mathcal V}_{n,k}}} \right|}}}\sum\nolimits_{i \in {{\cal V}_{n,k}}} \mathbb{E}{\left\{ {{d_{i,l}}\left| {k\Delta t \le t_{i,l}^{\rm{a}} < \left( {k + 1} \right)\Delta t } \right.} \right\}}, n \in \mathcal{N}.\notag
\end{equation}
\end{small} {Meanwhile, the average Packet Drop Ratio (PDR) of UE quantifies the communication reliability, which should be considered as another metric of QoS. It is defined as},\begin{small}
\begin{equation}
    \bar {\beta}_{n,k}{\rm{ = }}\frac{1}{{\left| {{{\mathcal V}_{n,k}}} \right|}}\sum\nolimits_{i \in {{\cal V}_{n,k}}} \mathbb{E}{\left\{ {L_{i,l}\left| {k\Delta t \le t_{i,l}^{\rm{a}} < \left( {k + 1} \right)\Delta t } \right.} \right\}}, n \in \mathcal{N}.\notag
\end{equation}
\end{small}Then, we define the reward function for slice $n$ at the $k$-th epoch as\begin{small}
\begin{align}
    &{J_{n,k}} = \underbrace {{\alpha _{n,1}} \cdot U_{{\rm{QoS}}}^{{\rm{PDR}}}\left( {\bar {\beta}_{n,k},\bar {\beta}_n^{{\max}},\bar {\beta}_n^{\min }} \right)}_{{\rm{PDR\;\;related\;\;reward}}}  \notag \\
    &\;\;\;\;\;\;\;\;\;\;\;\;\;\;\;\;\;\;\;\;\;\;\;\;\;\;\;\;\;\;\;\;\;\;
    + \underbrace {{\alpha _{n,{\rm{2}}}} \cdot U_{{\rm{QoS}}}^{{\rm{Lat}}}\left( {{{\bar d}_{n,k}},\bar d_n^{{\max}},\bar d_n^{\min }} \right)}_{{\rm{Packet\;\;Delay\;\;related\;\;reward}}},
    \label{eq_7} 
\end{align}
\end{small}where \begin{small}$U_{{\rm{QoS}}}^{{\rm{PDR}}}\left(  \cdot  \right)$\end{small} is a normalized reward function of average PDR \begin{small}$\bar {\beta}_{n,k}$\end{small} of slice $n$. To stabilize the learning procedure of the proposed DRL algorithm  developed in Section IV, \begin{small}$U_{{\rm{QoS}}}^{{\rm{PDR}}}\left(  \cdot  \right)$\end{small} is designed as a piecewise-linear concave function in (\ref{eq_7a}), where \begin{small}$\bar {\beta}_n^{{\rm{min}}}$\end{small} and \begin{small}$\bar {\beta}_n^{\max }$\end{small}  {denotes the min (target) and maximum tolerant PDR} for V2V service in for slice $n$. ${\alpha}_{n,1}$ is the maximum revenue, when \begin{small}$\bar {\beta}_{n,k}$\end{small} is less than target PDR value \begin{small}$\bar {\beta}_n^{{\rm{min}}}$\end{small}.
\par
\begin{figure*}[ht]
\begin{footnotesize}
\begin{numcases}
{U_{{\rm{QoS}}}^{{\rm{PDR}}}\left( {\bar {\beta}_{n,k},\bar {\beta}_n^{{\max}},\bar {\beta}_n^{\min }} \right) =}
{1,\;\;\bar {\beta}_n^{\min } > \bar {\beta}_{n,k} \ge 0, \notag} \\
{{{\left( {\bar \beta _n^{\max } - {{\bar \beta }_{n,k}}} \right)} \mathord{\left/
 {\vphantom {{\left( {\bar \beta _n^{\max } - {{\bar \beta }_{n,k}}} \right)} {\left( {\bar \beta _n^{\max } - \bar \beta _n^{\min }} \right)}}\label{eq_7a}} \right.
 \kern-\nulldelimiterspace} {\left( {\bar \beta _n^{\max } - \bar \beta _n^{\min }} \right)}},\;\;\bar {\beta}_n^{{\rm{max}}}\; > \bar {\beta}_{n,k} \ge \bar {\beta}^{\min}, }  \\
 {0,\;\;\bar {\beta}_{n,k} > \bar {\beta}_n^{\max},} \notag
\end{numcases}
\end{footnotesize}
\begin{footnotesize}
\begin{numcases}
{U_{{\rm{QoS}}}^{{\rm{Lat}}}\left( {{{\bar d}_{n,k}},\bar d_n^{{\max}},\bar d_n^{\min }} \right) =}
{1,\;\;\bar d_n^{{\rm{min}}} > {{\bar d}_{n,k}} > 0, \notag } \\
{{{\left( {\bar d_n^{\max } - {{\bar d}_{n,k}}} \right)} \mathord{\left/
 {\vphantom {{\left( {\bar d_n^{\max } - {{\bar d}_{n,k}}} \right)} {\left( {\bar d_n^{\max } - \bar d_n^{\min }} \right)}}} \right.
 \kern-\nulldelimiterspace} {\left( {\bar d_n^{\max } - \bar d_n^{\min }} \right)}},\;\;\bar d_n^{\max } > {{\bar d}_{n,k}} \ge \bar d_n^{\min },\label{eq_7b}} \\
{0,\;\;{{\bar d}_{n,k}} \ge \bar d_n^{\max }, \notag}
\end{numcases}
\end{footnotesize}
\hrulefill
\end{figure*}
Meanwhile, \begin{small}$U_{{\rm{QoS}}}^{{\rm{Lat}}}\left(  \cdot  \right)$\end{small} in (\ref{eq_7b}) is a normalized reward function of average packet delay ${\bar d}_{n,k}$ of slice $n$, where \begin{small}$\bar d_n^{{\max}}$\end{small} and \begin{small}$\bar d_n^{\min}$\end{small} denote the  {maximum tolerant value and minimum (target)} of packet delay for V2V service in slice $n$. $\alpha_{n,2}$ is the maximum revenue, when \begin{small}${\bar d}_{n,k}$\end{small} is less than minimum packet delay \begin{small}$\bar d_n^{{\max}}$\end{small}. 
\par
Therefore, at each epoch $k$, the reward function of all slices can be defined as
\begin{small}
\begin{equation}
    J_{k}\left( {{{\boldsymbol O}_k}} \right) = \sum\nolimits_{n \in \mathcal{N}} {{J_{n,k}}}. 
    \label{eq_8}
\end{equation}
\end{small}The goal of this paper is to find the optimal network slicing configuration policy with weights ${\boldsymbol{\theta}^{*}}$ that can maximize the long-term reward of all slices, which can be formulated as 
\begin{small}
\begin{equation}
    \mathop {\max }\limits_{\boldsymbol \theta}  \left\{ {J\left( {{\pi _{\boldsymbol \theta} }} \right) = \mathbb{E}\left[ {\left. {\sum\nolimits_{k = 1}^\infty  {{\lambda ^{k - 1}}J_{k}\left( {{{\boldsymbol O}_k}} \right)} } \right|{\pi _{\boldsymbol \theta} }} \right]} \right\},
    \label{eq_9}
\end{equation}
\end{small}where $\lambda$ is the discount factor.
\subsection{PoMDP}
 {Since the network  slicing  configuration  policy $\pi$, defined in \textbf{\textit{Definition \ref{def4}}}, is based on the partial observation of network status, \begin{small}${\boldsymbol{O}}_k$\end{small}, instead of the complete network status.} Therefore, the formulated problem (\ref{eq_9}) can be treated as a PoMDP with an infinite horizon discounted reward. PoMDP is an extension of MDP by adding a set of observations and the corresponding observation model \cite{DRL_survey_1}, which is defined as follows,
\begin{itemize}
    \item \textbf{System State}: The complete network status at $k$-th epoch is denoted as \begin{small}$\boldsymbol{X}_k$\end{small}, which follows Markovian, but cannot be directly observed by eNB;
    \item \textbf{Observation}: At each epoch, the eNB (agent) indirectly observes the complete network status \begin{small}$\boldsymbol{X}_k$\end{small} through observation \begin{small}$\boldsymbol{O}_k$\end{small} in \textbf{\textit{Definition \ref{def1}}}, which can be seen as a stochastic function of \begin{small}$\boldsymbol{X}_k$\end{small}; 
    \item \textbf{Action}: The action of the controller is configuration of slices $\boldsymbol{C}_k$ in \textit{\textbf{Definition \ref{def4}}} and the discrete action space is $\mathcal{C}$ (the collection of candidate slice configuration);
    \item \textbf{Observation History}: The observation history at the $k$-th epoch $\boldsymbol{H}_k$, in \textbf{\textit{Definition \ref{def3}}};
    \item \textbf{Reward Function}: It is the revenue of all slice at each epoch \begin{small}$J_k(\boldsymbol{O}_k)$\end{small} in formula (\ref{eq_9});
    \item \textbf{Q Function}: It is expected long-term revenue from taking action $\boldsymbol{C}_k$ under observation history $\boldsymbol{H}_k$, which is
    \begin{small}
    \begin{align}
        &Q\left( {{{\boldsymbol H}_k},{{\boldsymbol C}_k}} \right) = 
        \mathbb{E}{_{\tau  > k}}\left[ {\sum\nolimits_{k' = k}^\infty  {{\lambda ^{k' - k}}{J_{k'}}\left( {{{\boldsymbol O}_{k'}}} \right)\left| {{{\boldsymbol{H}}_k},{{\boldsymbol{C}}_k}} \right.} } \right],
        \notag
    \end{align}
    \end{small}where $\tau  > k$ refers to the sampling trajectory of observations and actions after epoch $k$, 
    \begin{small}
    \[\tau  > k{\rm{ = }}\left( {{{\boldsymbol O}_k},{{\boldsymbol{C}}_k},{{\boldsymbol O}_{k + 1}},{{\boldsymbol{C}}_{k + 1}}, \cdots} \right).\]
    \end{small}
    \item \textbf{Value Function}: It represents the expected long-term revenue starting from observation history $\boldsymbol{H}_k$ and the relationship between Q function and value function is,
    \begin{small}
    \begin{equation}
        V\left( {{{\boldsymbol{H}}_k}} \right) = \sum\nolimits_{{\boldsymbol{C}} \in \mathcal{C}} {{\pi _\theta }\left( {{\boldsymbol{C}}\left| {{{\boldsymbol{H}}_k}} \right.} \right)}  \cdot Q\left( {{{\boldsymbol{H}}_k},{\boldsymbol{C}}} \right),
        \notag
    \end{equation}
    \end{small}where ${\pi_{\boldsymbol{\theta}} }\left( {{\boldsymbol{C}}\left| {{{\boldsymbol{H}}_k}} \right.} \right)$ is the probability of choosing configuration $\boldsymbol{C} \in \mathcal{C}$ at observation history $\boldsymbol{H}_k$, under stochastic policy $\pi_{\boldsymbol{\theta}}$.
\end{itemize}
\par
Unfortunately, PoMDP is a very difficult to solve in general and directly solving it suffers the high computational complexity.  {However, we need to emphasize that utilizing existing RL methods for the problem (\ref{eq_9}) will raise the following challenge.}
\begin{challenge}
\label{challenge1}
A key assumption underlying majority of RL algorithms is the full observability of system status. However, in this paper, only a  {partial observation of vehicular network status} is available for the eNB, which makes the existing RL methods inadequate.
\end{challenge}
\section{Solution based on Actor-Critic Deep Reinforcement Learning}
 {Deep Reinforcement Learning (DRL) can apply to a wide range of control problems}, since ANN can extract high-level features from raw input data and provide a good approximation of objective functions. Therefore, in this section, we develop a DRL algorithm that can  {deal with} PoMDP problem (\ref{eq_9}). Firstly, we briefly explain the principles of Actor-Critic based RL and show its potential for solving problem (\ref{eq_9}). Then, to deal with \textbf{\textit{Challenge \ref{challenge1}}}, we propose an Actor-Critic DRL algorithm, which can obtain the optimal slicing configuration policy $\pi_{{\boldsymbol \theta}^{*}}$ from the observation history of vehicular networks, without requiring prior expert knowledge of networks.
\subsection{Actor-Critic based RL for Solving the Formulated Problem}

Generally, there are two categories of RL methods: 1) (e.g. Q-learning) and 2) RL based on policy search (e.g. policy gradient) \cite{DRL_survey_1}. Under the basic assumption of Markovian property, the value-based RL methods construct a value/Q function model for estimating how good each state, or state-action pair is, and then search for the optimal policy implicitly by optimizing the value/Q function. The RL based on value function have a good sampling efficiency and stable performance, but at cost of introducing bias in estimating of the value/Q function. On the other hand, without maintaining a value function model, the policy search methods directly search for the optimal policy by the approximated gradient with respect to the parameters of policy. Compared to the value-based RL methods, policy search methods can obtain a good policy with a faster convergence rate, which can be extended to the non-Markovian scenarios (e.g. PoMDP problems). However, this category usually tends to converge to a local optimal and suffer from higher variance and lower sample efficiency.
\par
To deal with these disadvantages, the Actor-Critic method, a hybrid of both policy-based and value-based method, is proposed. Particularly, as comparison of the value-based methods and the policy-based methods, we highlight two key advantages of the actor-critic methods in the following:
\begin{itemize}
    \item The actor-critic methods can be applied for non-Markovian scenario, such as the PoMDP problem (\ref{eq_9}), where only the observation of vehicular networks is available at the controller;
    \item It can balance the trade-off between the variance of policy gradient and bias of value function estimation, as well as the satisfactory convergence property.
\end{itemize}
Thus, we utilize the actor-critic method to solve problem (\ref{eq_9}). The “actor part” updates the policy in the direction given by  {the “critic part”}, that is,
\begin{enumerate}[a).]
\item \textbf{The Actor Part}: It uses the policy gradient method to search the best performing policy over a set of parametrized policies $\pi_{\boldsymbol{\theta}}$, where vector $\boldsymbol{\theta}$ is the parameters of RAN slicing policy $\pi$ defined in \textbf{\textit{Definition \ref{def4}}}. It is assumed that the policy $\pi_{\boldsymbol{\theta}}$ is differentiable with respect to parameter vector $\boldsymbol{\theta}$, and the gradient of the objective function \begin{small}$J(\boldsymbol{\theta})$\end{small} in the problem (\ref{eq_9}) is denoted as \begin{small}${\nabla_{\boldsymbol{\theta}}}J\left( {{\pi_{\boldsymbol{\theta}}}} \right)$\end{small}. Then, the maximum of the objective function \begin{small}$J(\boldsymbol{\theta})$\end{small} can be obtained by ascending the gradient of the objective function \begin{small}${\nabla_{\boldsymbol{\theta}}}J\left( {{\pi_{\boldsymbol{\theta}}}} \right)$\end{small}. The policy gradient update for the parameter vector $\boldsymbol{\theta}$ is given by\begin{small}
    \begin{equation}
        \boldsymbol{\theta}  \leftarrow \boldsymbol{\theta} + \eta  \cdot {\nabla _{\boldsymbol{\theta}} }J\left( {{\pi _{\boldsymbol{\theta}} }} \right)
        \label{eq_12}
    \end{equation}
    \end{small}where $\eta > 0$ is the learning rate for the policy update.
    \item \textbf{The Critic Part}: According a value function estimation model, the goal of the critic part is to evaluate the performance of the policy $\pi_{\boldsymbol{\theta}}$ and use it to calculate \begin{small}$\nabla_{\boldsymbol{\theta}}J{\left( {\pi_{\boldsymbol{\theta}} } \right)}$\end{small}. For problem (\ref{eq_9}), we can  {design an ANN} to approximate the value function and  {update the weights} of ANN utilizing the observation data set of vehicular networks.
\end{enumerate}
Therefore, we aim at designing a mode-free DRL algorithm with the actor-critic structure to solve problem (\ref{eq_9}). However, two main technical challenges arise as follows:
\begin{itemize}
    \item \textbf{Challenge 2}: How to deduce the policy gradient \begin{small}${\nabla_{\boldsymbol{\theta}} }J\left( {{\pi_{\boldsymbol{\theta}}}} \right)$\end{small} for the PoMDP problem (\ref{eq_9})? Since existing policy gradient methods can only apply to the Markovian scenario.
    \item \textbf{Challenge 3}: The input of the RAN slicing policy $\pi_{\boldsymbol{\theta}}$ is observation history $\boldsymbol{H}_k$, which is a time sequence. How to perform temporal abstraction of $\boldsymbol{H}_k$ in a saleable way?
\end{itemize}
\subsection{Proposed DRL Algorithm with Actor-Critic Structure}
Firstly, we start with the policy gradient \begin{small}${{\nabla}_{\boldsymbol{\theta}} }J\left( {{{\pi}_{\boldsymbol{\theta}}}} \right)$\end{small} to deal with \textbf{\textit{Challenge 2}}. It is noteworthy that the detailed design of DRL algorithm will be discussed  {later}. Here, we propose the customized policy gradient for PoMDP.
\\
\begin{theorem}[\textbf{Proposed Advantage Actor-Critic DRL algorithm  for PoMDP Problem}]
Following the idea of Advantage Actor-Critic (A2C) methods for the MDP problems, the policy gradient \begin{small}${{\nabla}_{\boldsymbol{\theta}} }J\left( {{{\pi}_{\boldsymbol{\theta}}}} \right)$\end{small} for the formulated PoMDP problem (\ref{eq_9}) is given by,\begin{small}
\begin{align}
    & {\nabla _{\boldsymbol{\theta}} }J\left( {{\pi _{\boldsymbol{\theta}} }} \right) = \notag\\
    &\;\;\;\;\;\;\;\;\mathbb{E}{_\tau }\left[ {\sum\limits_{k = 1}^\infty  {{\lambda ^{k - 1}} \cdot {\nabla _{\boldsymbol{\theta}} }\log {\pi _{\boldsymbol{\theta}} }\left( {{{\boldsymbol{C}}_k}\left| {{{\boldsymbol{H}}_k}} \right.} \right) \cdot A\left( {{{\boldsymbol{H}}_k},{{\boldsymbol{C}}_k}} \right)} } \right],
    \label{eq_13}
\end{align}
\end{small}where \begin{small}$\tau = ({\boldsymbol{O}_1},{\boldsymbol{C}_1},{\boldsymbol{O}_2},...)$\end{small} is  {a trajectory} of network status, and function \begin{small}$A({\boldsymbol{H}_k},{\boldsymbol{C}_k})$\end{small} is the advantage function, which is\begin{small}
\begin{equation}
    A\left( {{{\boldsymbol{H}}_k},{{\boldsymbol{C}}_k}} \right) = Q\left( {{{\boldsymbol{H}}_k},{{\boldsymbol{C}}_k}} \right) - V\left( {{{\boldsymbol{H}}_k}} \right). \tag{\ref{eq_13}a}
\end{equation}
\end{small}
\end{theorem}
\begin{IEEEproof}
    Detailed derivations are given in \textbf{Appendix A}.
\end{IEEEproof}

\begin{remark}
\textbf{Theorem 1} acquires an explicit form of policy gradient \begin{small}${\nabla _{\boldsymbol{\theta}} }J\left( {{\pi _{\boldsymbol{\theta}} }} \right)$\end{small}, where the policy gradient (\ref{eq_13}) is an expectation over entire trajectory $\tau$ of vehicular networks. We can compute gradient (\ref{eq_13}) approximately by using Monte-Carlo estimation. Specifically, the paradigm of policy gradient in (\ref{eq_13}) is like the concept to the maximum likelihood (ML) approaches in supervised learning, except that the policy gradient is weighted by sums of advantage functions over the trajectory. In fact, these sums of advantage functions may be positive or negative, thus the policy gradient will try to decrease the likelihood of samples with ``negative sums of advantage functions” and increase the likelihood of others. 
\end{remark}
\par
Based on \textbf{Theorem 1}, we propose a DRL algorithm  with actor-critic structure. In the this framework, the critic part utilizes an ANN to approximate the advantage function \begin{small}$A(\boldsymbol{H}_k,\boldsymbol{C}_k)$\end{small}, which is defined in formula (\ref{eq_13}a), while the actor part utilizes the gradient formula (\ref{eq_13}) to estimate the policy gradient \begin{small}${\nabla _{\boldsymbol{\theta}} }J\left( {{\pi _{\boldsymbol{\theta}} }} \right)$\end{small} and then updates the parameter vector $\boldsymbol{\theta}$ of the RAN slicing policy $\pi_{\boldsymbol{\theta}}$ according to the formula (\ref{eq_12}) in Section \ref{3-1}.
\subsection{Implementation of DRL with Actor-Critic Structure}
\subsubsection{Critic Part} Compared to the Q-function \begin{small}$Q(\boldsymbol{H}_k,\boldsymbol{C}_k)$\end{small} and the advantage function \begin{small}$A(\boldsymbol{H}_k,\boldsymbol{C}_k)$\end{small}, the value function \begin{small}$V(\boldsymbol{H}_k)$\end{small} is the simplest one since it only depends on the observation history \begin{small}$\boldsymbol{H}_k$\end{small} and thus is hoped to be easier for the critic part to learn.  {With the value function \begin{small}$V(\boldsymbol{H}_k)$\end{small}, the Q-function \begin{small}$Q(\boldsymbol{H}_k,\boldsymbol{C}_k)$\end{small} can be approximated by sample value. Then, the advantage function \begin{small}$A(\boldsymbol{H}_k,\boldsymbol{C}_k)$\end{small} can be approximated as follows}\begin{small}
\begin{equation}\label{A_approximate}
    {\hat A}\left( {{{\boldsymbol{H}}_k},{{\boldsymbol{C}}_k}} \right) = {\sum\nolimits_{k' = k}^\infty  {{\lambda ^{k' - k}}{J_{k'}}\left( {{{\boldsymbol O}_{k'}}} \right)}} - V\left( {{{\boldsymbol{H}}_k}} \right),
\end{equation}
\end{small}where \begin{small}$\boldsymbol{O}_k$\end{small} and \begin{small}$\boldsymbol{H}_{k}$\end{small} are sampled through Monte-Carlo method.
\par
Therefore, the critic part approximates the value function \begin{small}$V(\boldsymbol{H}_k)$\end{small}, and then estimate the advantage function \begin{small}$A(\boldsymbol{H}_k,\boldsymbol{C}_k)$\end{small}. Like the idea of utilizing the Deep Q-Network (DQN) to fit the Q function, we introduce a critic neural network $\hat V$ to approximate the  {real value function \begin{small}$V(\boldsymbol{H}_k)$\end{small}}. However, as mentioned in \textbf{Challenge 2}, the critic neural network $\hat V$ is a function in terms of the observation history at the $k$-th epoch $\boldsymbol{H}_k$. To perform the temporal abstraction of observation history $\boldsymbol{H}_k$, we modify $\hat V$ by leveraging recent advances in recurrent neural networks (RNN), that is, replacing the first fully-connected layer with a Long Short-Term Memory (LSTM) layer, which is regarded as a memory cell with three different gates, which regulating the information and thus allowing to keep the past information \cite{R26}. Therefore, the LSTM layer can capture the longer-term temporal dependencies of observation history $\boldsymbol{H}_k$ as compared to the traditional RNNs.
\par

Therefore, the critic neural network \begin{small}$\hat V$\end{small} is represented as\begin{small}
\begin{equation}
    V\left( {{{\boldsymbol{H}}_k}} \right) \approx {\hat V_w}( {{\hat{\boldsymbol{H}}_k}}). \notag
\end{equation}
\end{small}where vector $\boldsymbol{w}$ is the weights of the critic neural network $\hat V$ and term\begin{small}
\begin{equation}
    {\hat{\boldsymbol H}_k} = \left( {{{\boldsymbol{O}}_{k-K}},...,{{\boldsymbol{O}}_{k - 1}}} \right). \notag
\end{equation}
\end{small}is a finite fixed-length window of past observations, which consists of the \begin{small}$K \in {\mathbb{N}}^{+}$\end{small} most recent observations of vehicular networks and actions to the $k$-th epoch.
\par
In the learning process of the critic neural network \begin{small}$\hat V$\end{small}, weights $\boldsymbol{w}$ is learned by minimizing the Mean Square Error (MSE) loss function at each learning step, which is given by
\begin{small}
\begin{equation}
    L\left( \boldsymbol{w} \right) = \frac{1}{2} \cdot \mathbb{E}\left\{ {{{\left[ {{{\hat V}_{\boldsymbol{w}}}( {{\hat{\boldsymbol{H}}_k}}) - {\sum\nolimits_{k' = k}^\infty  {{\lambda ^{k' - k}}{J_{k'}}\left( {{{\boldsymbol O}_{k'}}} \right)}}} \right]}^2}} \right\}.
    \label{eq_22}
\end{equation}
\end{small}Then, the weights $\boldsymbol{w}$ of the critic neural network \begin{small}${\hat V_{{\boldsymbol{w}}}}( {{\hat{\boldsymbol{H}}_k}} )$\end{small} are updated as\begin{small}
\[\boldsymbol{w} \leftarrow {\boldsymbol{w}}{\rm{ + }}\nu  \cdot {{\nabla}_{\boldsymbol{w}}}L\left( \boldsymbol{w} \right),\]
\end{small}where $\nu$ is the learning rate for weights $\boldsymbol{w}$.
\par
In the following, we describe the detailed structure of the critic neural network \begin{small}${\hat V}_{\boldsymbol{w}}({\hat {\boldsymbol{H}}}_k)$\end{small}. 
\begin{enumerate}[a).]
    \item \textit{Input}: The input \begin{small}$\boldsymbol{O}_k$\end{small} (i.e. observation of vehicular networks) to the critic neural network is  {a vector size $2N$, where the $2n+1$-th to $2(n+1)$-th input entries corresponds to the observation of slice $n \in \mathcal{N}$ at the $k$-th epoch, \begin{small}$\boldsymbol{O}_{n,k}$\end{small}}.
    \item \textit{LSTM layer}: It maintains an internal state and aggregate observation states over time. This gives the critic neural network is responsible of learning how to aggregate observation states over time. We use the Rectified Linear Unit (ReLU) as the activation function for the LSTM layer.
    \item \textit{Hidden layers}: The number of neurons of each hidden layer is the same, and ReLU function is used as the activation function.
    \item \textit{Output layer}: The output of the DQN is a scalar, which is the estimated value of function \begin{small}$V(\boldsymbol{H}_k)$\end{small} under current observation history $\boldsymbol{H}_k$. 
\end{enumerate}
\subsubsection{Actor Part}
Based on \textbf{Theorem 1} and policy gradient update equation (\ref{eq_12}), the weights \begin{small}$\boldsymbol{\theta}$\end{small} of RAN slicing policy $\pi_{\boldsymbol{\theta}}$ are updated as\begin{small}
\begin{equation}
    \boldsymbol{\theta}  \leftarrow \boldsymbol{\theta}  + \eta  \cdot {\nabla _{\boldsymbol{\theta}} }J\left( {{\pi _{\boldsymbol{\theta}} }} \right) \notag
\end{equation}
\end{small}where \begin{small}$\eta \in {\mathbb{R}^{+}}$\end{small} is the learning rate for the update of parameter $\boldsymbol{w}$ and\begin{small}
\[{\nabla _{\boldsymbol{\theta}} }J\left( {{\pi _{\boldsymbol{\theta}} }} \right) \approx \sum\nolimits_k {\left[ {{\lambda ^{k - 1}} \cdot {\nabla _{\boldsymbol{\theta}} }\log {\pi _{\boldsymbol{\theta}} }\left( {{{\boldsymbol{C}}_k}\left| {{{\boldsymbol{H}}_k}} \right.} \right) \cdot \hat A\left( {{{\boldsymbol{H}}_k},{{\boldsymbol{C}}_k}} \right)} \right]} ,\]
\end{small}and the advantage function \begin{small}$\hat A({{{\boldsymbol{H}}_k},{{\boldsymbol{C}}_k}})$\end{small} is defined in (\ref{A_approximate}).
\par
Furthermore, to evaluate the training performance of the actor neural network $\pi_{\boldsymbol{\theta}}$, we define the loss function of the actor neural network as\begin{small}
\begin{equation}
    L\left( {\boldsymbol{\theta}}  \right) = \sum\nolimits_k {\left[ {{\lambda ^{k - 1}} \cdot {\nabla _{\boldsymbol{\theta}} }\log {\pi _{\boldsymbol{\theta}} }\left( {{{\boldsymbol{C}}_k}\left| {{{\boldsymbol{H}}_k}} \right.} \right) \cdot \hat A\left( {{{\boldsymbol{H}}_k},{{\boldsymbol{C}}_k}} \right)} \right]}.
    \label{eq_25}
\end{equation}
\end{small}\par
 {Actor neural network $\pi_{\boldsymbol{\theta}}$ has the same structure as the critic neural network \begin{small}$\hat V_{\boldsymbol{w}}$\end{small}. It is noteworthy that the actor neural network and the critic neural network share the same LSTM layer, as illustrated in Figure 3. This setting can make the actor and critic neural network shares the same hidden states of LSTM layer, which makes DRL training more stable.} The output layer of policy $\pi_{\boldsymbol{\theta}}$ is described as follows: the output of policy $\pi_{\boldsymbol{\theta}}$ is a vector of size \begin{small}$\left|  \mathcal{C}\right|$\end{small}, where each element of the output layer is mapping to the probability of candidate configuration \begin{small}${\boldsymbol{C}} \in {\mathcal C}$\end{small} under current observation history \begin{small}${\hat{\boldsymbol{H}}_k}$\end{small}.\begin{figure}[!h]
	\centering
	\includegraphics[scale=0.28]{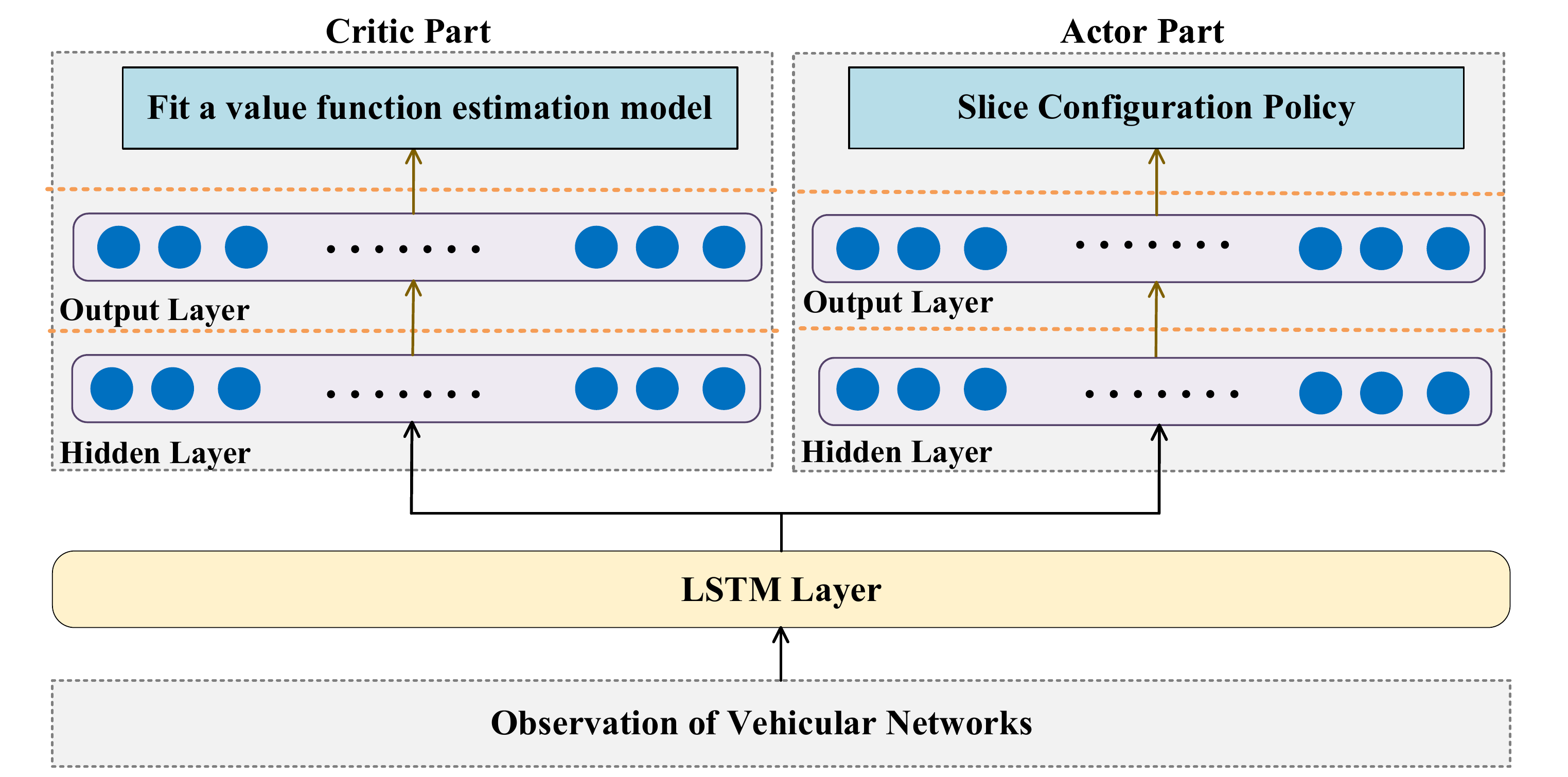}  
	\caption{ {The structure of the critic neural network and actor neural network.}}
	\label{Fig5}
\end{figure}
\par
The overall learning procedure of our proposed advantage actor-critic DRL algorithm  is provided in \textbf{Algorithm \ref{alg1}}.
\begin{figure}[!h]
	\begin{MYalgorithmic}
		\algcaption{Training of the RAN slicing policy $\pi_{\boldsymbol{\theta}}$}
		\label{alg1}
	\begin{footnotesize}
	\begin{algorithmic}
		\renewcommand{\algorithmicrequire}{\textbf{Initialization:}}
		\Require
		\State Initialize the critic neural network $\hat{V}$ and the RAN slicing policy (i.e., the actor part) $\pi_{\boldsymbol{\theta}}$ with weights $\boldsymbol{w}$ and $\boldsymbol{\theta}$. Initialize the replay buffer $R$. Initialize the length of observation history $K$ and the empty observation history.
		\renewcommand{\algorithmicrequire}{\textbf{Repeat:}}
		\Require
		\State \textbf{1)} Receive observation of vehicular networks \begin{small}$\boldsymbol{O}_k$\end{small}.
		\State \textbf{2)} Append observation and previous slice configuration to history, ${\hat {\boldsymbol{H}}}_k \leftarrow ({\hat {\boldsymbol{H}}}_{k-1},\boldsymbol{O}_{k-1})$.
		\State \textbf{3)} Select slice configuration, $\boldsymbol{C}_k \leftarrow \pi_{\boldsymbol{\theta}}(\boldsymbol{H}_k)$.
		\State \textbf{4)} Store the sample trajectory $\left\{(\boldsymbol{H}_k,\boldsymbol{C}_k):k=1,...,T\right\}$.
	    \State \textbf{5)} Update the critic neural network ${\hat{V}_{\boldsymbol{w}}}$ using equation (\ref{eq_22}),
	    \[{\boldsymbol{w}} \leftarrow {\boldsymbol{w}}{\rm{ + }}\nu  \cdot {\nabla _{\boldsymbol{w}}}L\left( {\boldsymbol{w}} \right).\]
	    \State \textbf{6)} Update the RAN slicing policy (the actor part) $\pi_{\boldsymbol{\theta}}$ using equation (\ref{eq_25})
	    \[\boldsymbol{\theta}  \leftarrow \boldsymbol{\theta}  + \eta  \cdot {{\nabla}_{\boldsymbol{\theta}}}J\left( {{{\pi} _{\boldsymbol{\theta}} }} \right)\]
		\Require
 	\end{algorithmic}
	\end{footnotesize}
\end{MYalgorithmic}
\end{figure}
\par
\subsubsection{Complexity of the Proposed DRL algorithm }
 {
According to analysis method in~\cite{Complexity}, the computational complexity of learning procedure for the proposed DRL algorithm (i.e., \textbf{Algorithm 1}) can be expressed by\begin{small}
\begin{equation}
   {\cal O}\left( {T_{\rm{L}} \cdot \left( {\sum\nolimits_{l = 0}^{{L_{{\rm{actor}}}}} {n_{{\rm{actor}}}^{\left( l \right)} \cdot n_{{\rm{actor}}}^{\left( {l + 1} \right)}}  + \sum\nolimits_{l = 0}^{{L_{{\rm{critic}}}}} {n_{{\rm{critic}}}^{\left( l \right)} \cdot n_{{\rm{critic}}}^{\left( {l + 1} \right)}} } \right)} \right), \notag
\end{equation}
\end{small}where \begin{small}$T_{\rm{L}}$\end{small} is the learning steps of slicing configuration policy training, \begin{small}$n_{{\rm{actor}}}^{\left( l \right)}$\end{small} is the number of neurons in the $l$-th layer of the actor part, i.e., neural network \begin{small}${\pi _{\boldsymbol{\theta}}}$\end{small}, \begin{small}$n_{{\rm{critic}}}^{\left( l \right)}$\end{small} is the number of neurons in the $l$-th layer of the critic neural network, i.e., neural network \begin{small}${\hat V_w}$\end{small}, and \begin{small}${L_{{\rm{actor}}}}\;\left({L_{{\rm{critic}}}}\right)$\end{small} denotes the number of the hidden layers in the actor part (critic part).}

\section{Simulation Results and Analysis}
In order to demonstrate the effectiveness of our proposed network slice self-configuration scheme, a system level simulation platform is implemented. Herein, we consider a six-lane freeway and each direction has three lanes, where the length of the freeway is 3.4 $km$ and the width of lane is set as 4 m (A 1.2, Annex A, 3GPP 36.885 \cite{3gpp01}). Software including MATLAB 2019a and Keras 2.2.2 with Python 3.5.2. are utilized for simulations. There are two types of services and two corresponding slices are considered in the simulation: a) network slice for traffic safety related service, which aims at reducing the possibility of traffic accidents and improvement of traffic efficiency; b). network slice for autonomous driving related service, which is utilized for the cooperative awareness and control between autonomous vehicles. Since the critical nature of communication reliability in V2V services, we set higher weighting factor for the PDR related function $U_{\rm QoS}^{\rm PDR}$ in the reward function (\ref{eq_7}). The discount rate $\lambda$ in the formulated problem (\ref{eq_9}) for estimating long-term reward function is 0.9.
\par
Based on the observation of vehicular networks \begin{small}$\boldsymbol{O}_k$\end{small}, the proposed DRL algorithm  trains the RAN slicing policy by \textbf{Algorithm 1}. Furthermore, in \textbf{Algorithm 1}, the critic neural network $\hat V$ is a four layers neural network. The LSTM layer contains 256 units and uses Rectified Linear Unit (ReLU) as the activation function. There is one hidden layers in $\hat V$. The hidden layer contains 64 units and uses ReLU as the activation function. With linear activation function, the output layer gives the estimated value function. On the other hand, the actor neural network $\pi_{\boldsymbol{\theta}}$ is has same structure to the critic neural network $\hat V$, which also has one hidden layer. Besides, the critic neural network $\hat V$ and the actor neural network $\pi_{\boldsymbol{\theta}}$ are learned with a learning rate of $10^{-4}$. 
\begin{table*}[!t]
	\centering
	\newcommand{\tabincell}[2]{\begin{tabular}{@{}#1@{}}#2\end{tabular}}
	\footnotesize
	\renewcommand{\arraystretch}{1.1}
	\caption{Default Parameter settings for simulation.}
	\label{table2}
	{\begin{tabular}{|l|l|l|l|l|l|l|}
		\hline
		\multicolumn{2}{|l|}{\textbf{Parameter}} & \multicolumn{2}{|l|}{\textbf{Assumption}} \\
		\hline
		\multicolumn{2}{|l|}{Carrier frequency/Bandwidth/Number of RBs} & \multicolumn{2}{|l|}{5.9 GHz/ 10 MHz/ 50} \\
		\hline
		\multicolumn{2}{|l|}{Pathloss model/Small-scale fading} & \multicolumn{2}{|l|}{WINNER+ B1/  {Rician fading}}\\
		\hline
		\multicolumn{2}{|l|}{Total Transmit Power of VUE} & \multicolumn{2}{|l|}{20 dBm} \\
		\hline
		\multicolumn{2}{|l|}{The length of each epoch} & \multicolumn{2}{|l|}{400 $ms$ (i.e., 400 slot)} \\
		\hline%
		\multicolumn{2}{|l|}{Absolute vehicle speed} & \multicolumn{2}{|l|}{70 km/h} \\
		\hline
		\multicolumn{2}{|l|}{Average number of VUEs in the vehicular network} & \multicolumn{2}{|l|}{100} \\
		\hline
		\multicolumn{2}{|l|}{Service type} & \multicolumn{1}{|l|}{Traffic safety related service} & \multicolumn{1}{|l|}{Autonomous driving related service} \\
		\hline
		\multicolumn{2}{|l|}{Weighting factors in utility function (\ref{eq_7})}  & \multicolumn{1}{|l|}{$\boldsymbol{\alpha}_1=\left[1,2\right]$} & \multicolumn{1}{|l|}{$\boldsymbol{\alpha}_2=\left[1,3\right]$} \\
		\hline
		\multicolumn{2}{|l|}{Packet size per UE} & \multicolumn{1}{|l|}{300 Byte} & \multicolumn{1}{|l|}{200 Byte} \\
		\hline
		\multicolumn{2}{|l|}{Packet arrival period} & \multicolumn{1}{|l|}{50 $ms$} & \multicolumn{1}{|l|}{25 $ms$}\\
		\hline
		\multirow{3}{*}{Candidate slicing configurations} & {$F_{n}$} & {2 or 3 or 4} & {2 or 3 or 4}\\
		\cline{2-4}
        & {$B_{n}$} & {1.44 or 2.16 MHz} & {1.08 or 1.44 MHz}\\
        \cline{2-4}
        & {$T_{n}^{\rm{sw}}$} & {30 $ms$ or 50 $ms$} & {25 $ms$ or 15 $ms$} \\
		\hline
		 \multicolumn{2}{|l|}{ {The number of candidate slicing configurations}} & \multicolumn{2}{|l|}{ {36}}  \\
		\hline
	\end{tabular}}
\end{table*}

\subsection{Training Performances}

	\begin{figure}[!h]
		\centering
		\subfigure[Training Loss of Critic neural network]{
			\begin{minipage}[b]{0.96\linewidth}
				\centering
				\includegraphics[width=1\linewidth]{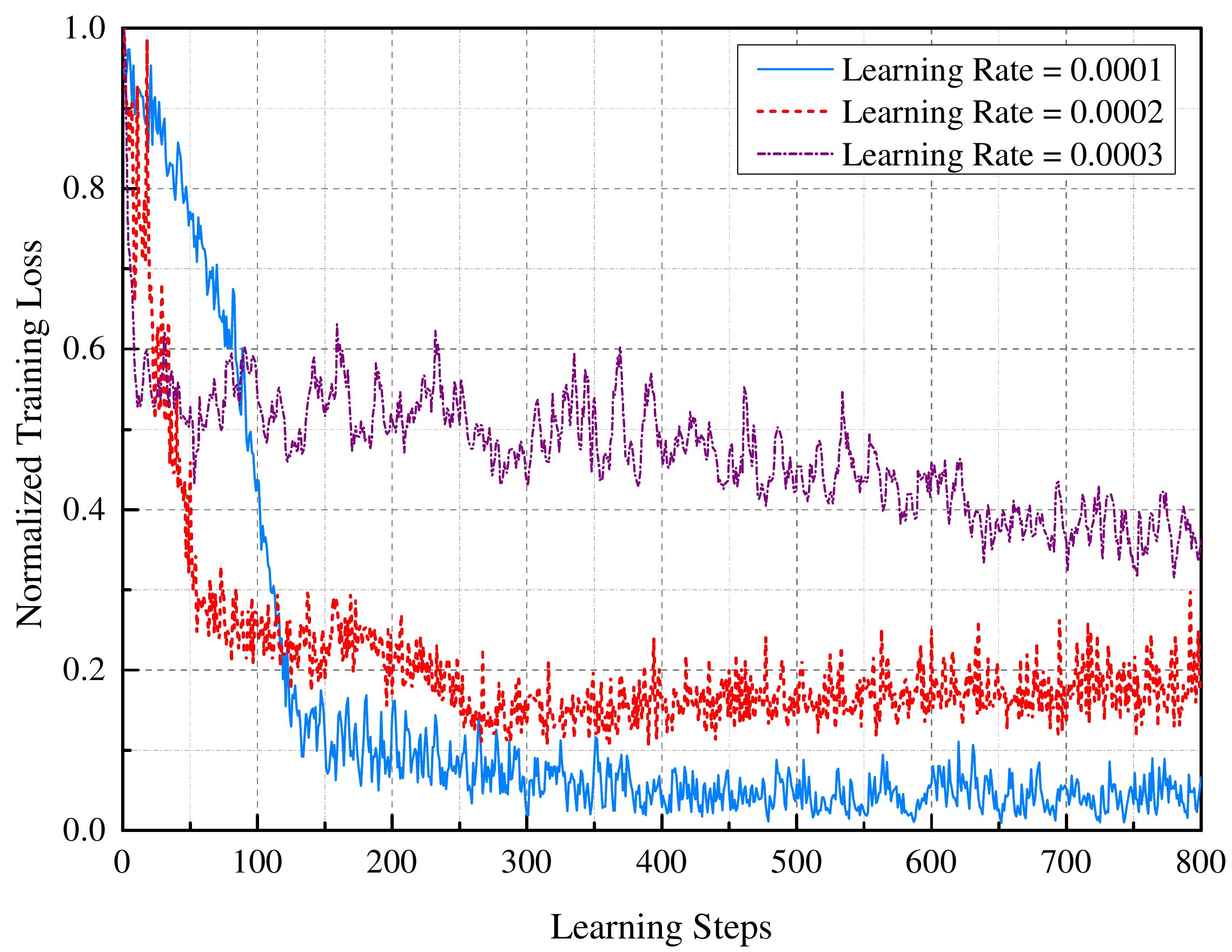}
			\end{minipage}
		}
		
		\subfigure[Training Loss of Actor neural network]{
			\begin{minipage}[b]{0.96\linewidth}
				\centering
				\includegraphics[width=1\linewidth]{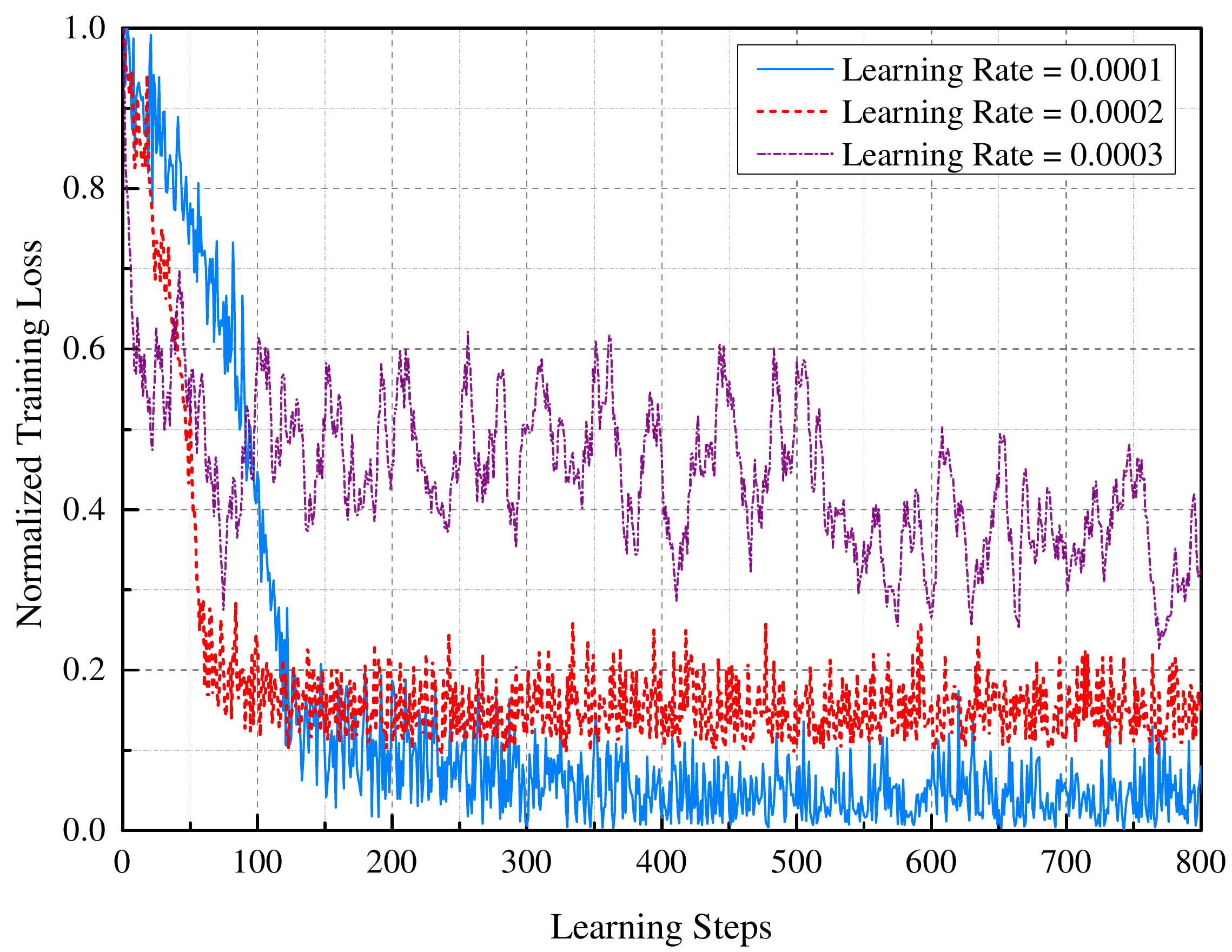}
			\end{minipage}
		}
		\caption{ {Illustration of the convergence of the proposed DRL algorithm under different learning rates}: 1) the train loss of critic neural network ${\hat V}_{\boldsymbol{w}}$ is measured by loss function $L({\boldsymbol{w}})$ defined in (\ref{eq_22}); 2). the loss function of actor neural network $\pi_{\boldsymbol{\theta}}$ (i.e., the slice self-configuration policy) is estimated by loss function $L({\boldsymbol{\theta}})$ defined in (\ref{eq_25}).}
		\label{Fig3}
	\end{figure}
 
The first experiment aims to examine the convergence of the proposed actor-critic DRL algorithm. Under the default simulation setup, as shown in Figure \ref{Fig3},  {we plot the variations in the loss functions of the critic neural network ${\hat V}_{\boldsymbol{w}}$ and the actor neural network $\pi_{\boldsymbol{\theta}}$ under different settings of learning rate}. This metric can measure the convergence speed of the loss function during the training procedure in Algorithm 1. In Figure \ref{Fig3}(a),  {under default learning rate $10^{-4}$, the values of loss function $L({\boldsymbol{w}})$ decrease close to 0.02 after the 300 learning steps, which means the value function ${\hat V}_{\boldsymbol{w}}$ has reached to a stable performance. On the other hand, in Figure \ref{Fig3}(b), the values of loss function $L({\boldsymbol{\theta}})$ decrease quickly during the first 300 learning steps. After the first 300 learning steps, the value curve of the $L({\boldsymbol{\theta}})$ is convergent to the value of about 0.04, which indicates the slice self-configuration policy has evolved into convergence condition.} The results confirm that the proposed DRL algorithm can avoid the mis-convergence and unstable issues in the training procedure.
\par
 {
Since the learning rate is one crucial hyper-parameter in deep learning, we compare the learning trends under different learning rates. It can be seen that, with a higher value of learning rate, the convergence speed of both actor and critic neural networks is increasing, but at the cost of higher training loss and drastic fluctuation at the convergence condition. For instance, when learning rate equals to $3\times10^{-4}$, the train loss is ``stable" after 100 learning steps, but the value of train loss is dramatically is dramatically fluctuating. Therefore, there is a trade-off between convergence speed and convergence value of training loss.}
\subsection{Performance Evaluation}
\begin{figure}[!h]
	\centering
	\includegraphics[scale=0.36]{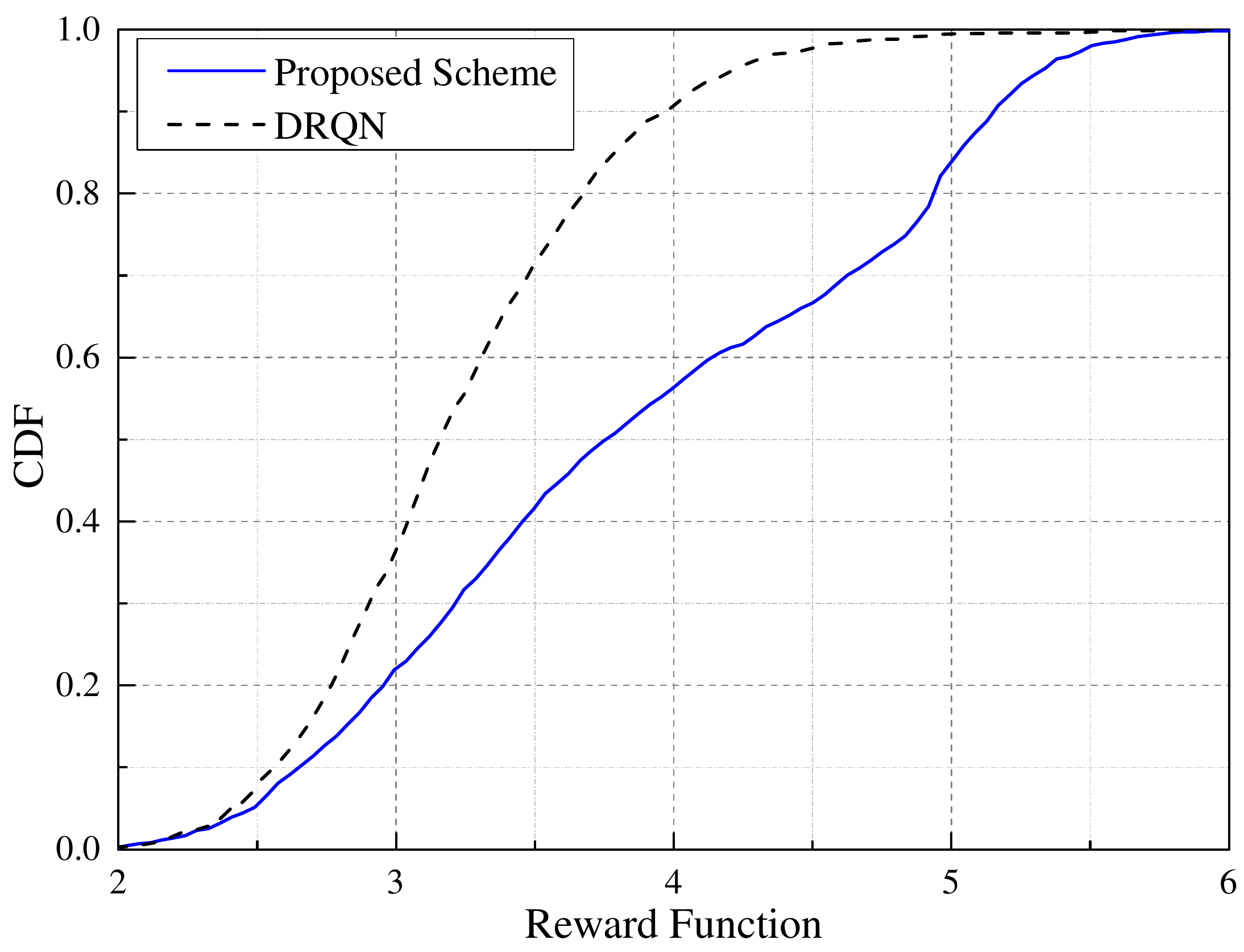}  
	\caption{ {CDF of the reward function in formula (\ref{eq_8}) with different schemes under default settings}: reward function $J$ can reflect the overall performance of network slicing control–the higher value of the function $J$, the better the QoS performance of network slices.}
	\label{Fig4}
\end{figure}
For the performance comparisons, we consider the Deep Recurrent Q-Network (DRQN), one of state-of-art DRL algorithms, as the baseline scheme \cite{Mei}. In the DRQN scheme, it replaces the first layer of the DQN with a LSTM layer with 256 units and ReLU activation function. In the simulation, DRQN is a four layers neural network. In the input LSTM layer, there is one input, i.e., the observation of vehicular networks \begin{small}$\boldsymbol{O}_k$\end{small}. Two hidden layers, each contains 128 units with the ReLU activation function. The output layer has the same size as the number of all candidate slicing configurations (i.e., $I_{\mathcal{C}}$), and each element of the output layer is mapping to an estimated value of \begin{small}$Q(\boldsymbol{O}_k,\boldsymbol{C}_k)$\end{small} under the observation \begin{small}$\boldsymbol{O}_k$\end{small}. In the baseline scheme, the DRQN is learned by using the Adam algorithm with a learning rate of $10^{-4}$, and weights of the target DRQN are copied from the weights of DRQN every 200 learning steps. Besides, ${\epsilon}$-greedy rule is used in the baseline scheme. The parameters of ${\epsilon}$-greedy rule are set as ${\epsilon}_{\rm min}=0.01$ and ${\epsilon}_{\rm decay}=0.01$. \par
From a statistical point of view, we evaluate the reward function defined in (8), which is used to indicate the QoS performance of slicing control. Figure \ref{Fig4} presents the cumulative distribution functions (CDFs) of the reward function with the proposed scheme and state-of-the-art DRQN scheme.  {It shows that the overall performance of the proposed scheme (mean value is 3.924) is better than the DRQN scheme (mean value is 
3.276). Specifically, the proposed DRL scheme can improve around 20\% than that of the DRQN scheme. The main reason is that, compared to the proposed DRL algorithm, the DRQN scheme, one kind of value-based DRL approaches, is introducing a higher level of bias in the estimation of the Q-value function, and then making the sub-optimal decisions. Then, the DRQN scheme mis-estimates the real status of network slices, and then make inappropriate adaption of slice configuration, which may deteriorate the PDR performance of each slice. In the following part, more experiment are carried out to show the RAN slicing performance of different schemes.}
\par
\begin{figure}[!t]
	\centering
	\subfigure[The slice for the autonomous driving related service.]{
		\begin{minipage}[b]{1\linewidth}
			\centering
			\includegraphics[width=1\linewidth]{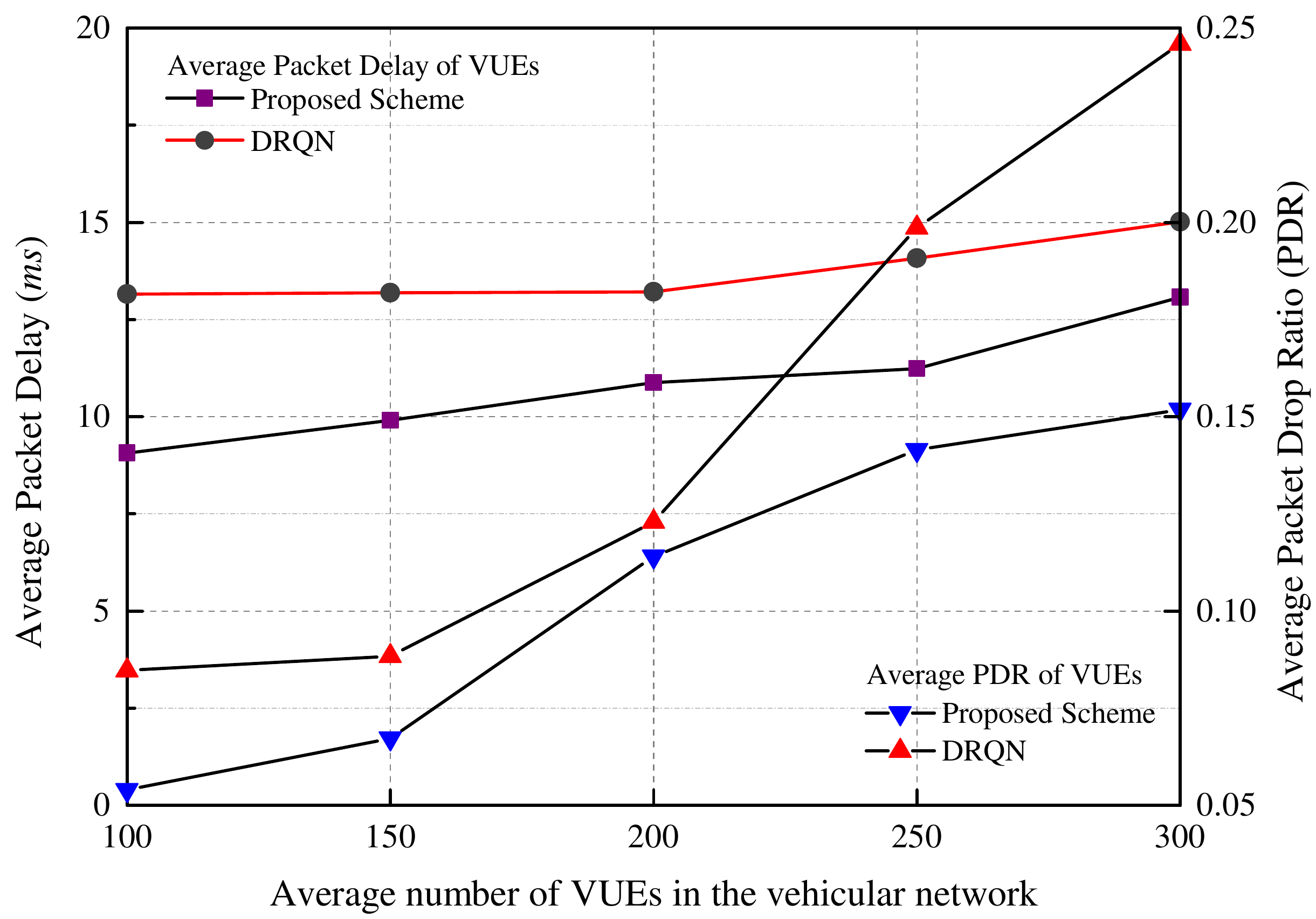}
		\end{minipage}
	}
	\subfigure[The slice for the traffic safety related service.]{
		\begin{minipage}[b]{1\linewidth}
			\centering
			\includegraphics[width=1\linewidth]{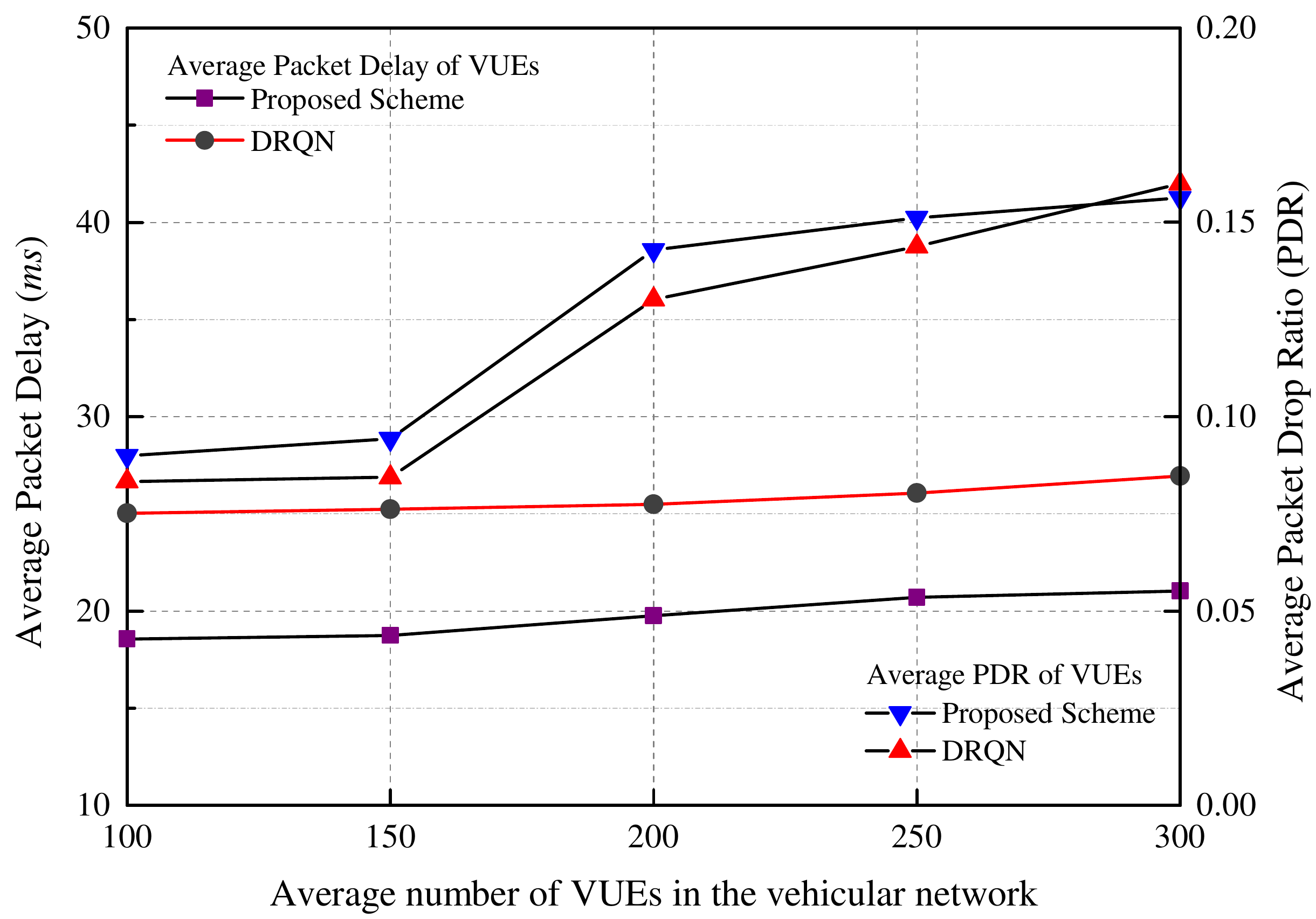}
		\end{minipage}
	}
	\caption{QoS performance of each network slice versus different vehicle density (i.e., the average number of VUEs in the network) under different schemes.}
	\label{Fig7}
	\vspace{-0.3cm}
\end{figure}
 {
As shown in the Figure \ref{Fig7}, it depicts the QoS performance of two considered network slices versus different vehicle density (i.e., the average number of VUEs in the network) under the DRQN and the proposed scheme. Figure \ref{Fig7}(a) depicts the QoS performance of UE in the slice for the autonomous driving related service. It can be seen that the average packet delay and PDR of UE decreases on the increasing vehicle density. Compared with the DRQN based network slicing, the proposed scheme has both lower average packet delay and PDR of UE. On the other hand, Figure \ref{Fig7}(b) depicts the QoS performance of UE in the slice for the traffic safety related service. It can be observed from Figure \ref{Fig7}(b) that, in the slice for the traffic safety related service, the proposed scheme has better average packet latency performance than the DRQN scheme. However, the DRQN scheme has slight better average PDR performance than the proposed scheme.}
\par
 { The main reason is that, in the reward function (\ref{eq_7}), the higher weighting factor for the PDR related to the slice for the autonomous driving related service, the proposed scheme tends to choose the slice configuration with more sub-channels and wider sub-channel bandwidth for the slice for the autonomous driving related service. This slicing strategy can ensure and improve the reward function but at cost of sacrificing the average PDR to a certain extent. The DRQN scheme is prone to choose the slice configuration with longer selection window length, which will decrease the average PDR but increase the average packet delay. Besides, when the vehicle density is increasing, the performance of DRQN is deteriorating rapidly. Overall, the result conforms to our expectations of the proposed DRL scheme for network slicing in C-V2X Mode 4 based networks.}
\section{Conclusion}
\label{sec:6}
In this paper,  {we propose an intelligent semi-decentralized network slicing framework for the C-V2X Mode 4 networks, which aims at maximizing the long-term QoS performance of V2V services}. Specifically, the proposed network slicing framework is implemented by a carefully designed actor-critic structured DRL algorithm.  {It has following advantages. Firstly, due to the proposed scheme has a semi-decentralize structure, the eNB only operates at a large timescale (in the level of hundreds of milliseconds), which has good scalability and can significantly reduce the signaling overhead. Meanwhile, because the eNB can infer the global view of vehicular networks from observation history, the decision-making process of the slice configuration at the eNB side can better ensure the control performance.} Simulation results show that the proposed scheme has stable convergence control performance and achieves higher QoS performance as compared to the state-of-art baseline scheme. However, due to the lack of data generated in real V2V service traffic, the traffic model is assumed as 3GPP model, which has a limitation in characterizing the realistic scenario. Last but not least, the proposed DRL algorithm is essentially an on-policy leaner, which has lower sample efficiency. These issues need to be further studied.

\appendix
\subsection{Proof of Theorem 1}
\begin{IEEEproof}
Firstly, based on the definition of value function in (\ref{eq_13}), the long-term revenue \begin{small}$J\left( {{\pi _{\boldsymbol{\theta}} }} \right)$\end{small} can be re-written as
\begin{small}
\begin{align}
J\left( {{\pi _{\boldsymbol{\theta}} }} \right) &= \mathbb{E}\left[ {\left. {\sum\nolimits_{k = 1}^\infty  {{\lambda ^{k - 1}}J\left( {{{\boldsymbol O}_k}} \right)} } \right|{\pi _{\boldsymbol{\theta}} }} \right] = \sum\nolimits_{{{\boldsymbol{H}}_1}} {\Pr \left[ {{{\boldsymbol{H}}_1}} \right]V\left( {{{\boldsymbol{H}}_1}} \right)} ,\notag
\end{align}
\end{small}where observation history \begin{small}$\boldsymbol{H}_1=(\boldsymbol{O}_1)$\end{small}. Then, gradient of \begin{small}$J\left( {{\pi _{\boldsymbol{\theta}} }} \right)$\end{small} can be written as
\begin{small}
\[{\nabla _{\boldsymbol{\theta}} }J\left( {{\pi _{\boldsymbol{\theta}} }} \right) = \sum\nolimits_{{{\boldsymbol{H}}_1}} {\Pr \left[ {{{\boldsymbol{H}}_1}} \right]{\nabla _{\boldsymbol{\theta}} }V\left( {{{\boldsymbol{H}}_1}} \right)} .\]
\end{small}Thus, our focus is to obtain the derivation of the value function, (i.e. \begin{small}${\nabla _{\boldsymbol{\theta}} }V\left( {{{\boldsymbol{H}}_k}} \right)$, $k=1,2,...$\end{small}),
\begin{small}
\begin{align}
&{{\nabla _{\bf{\theta }}}V\left( {{{\boldsymbol{H}}_k}} \right)} \notag \\
&={{\nabla _{\bf{\theta }}}\left( {\sum\nolimits_{{{\boldsymbol{C}}_k} \in {\mathcal{C}}} {{\pi _{\bf{\theta }}}\left( {{{\boldsymbol{C}}_k}\left| {{{\boldsymbol{H}}_k}} \right.} \right) \cdot Q\left( {{{\boldsymbol{H}}_k},{{\boldsymbol{C}}_k}} \right)} } \right)} \notag \\
&={\sum\nolimits_{{{\boldsymbol{C}}_k} \in {\mathcal{C}}} {\left[ {{{\nabla _{\bf{\theta }}}\left( {{\pi _{\bf{\theta }}}\left( {{{\boldsymbol{C}}_k}\left| {{{\boldsymbol{H}}_k}} \right.} \right)} \right) \cdot Q\left( {{{\boldsymbol{H}}_k},{{\boldsymbol{C}}_k}} \right)}} \right.} } \notag \\
&\;\;\;\;\;\;\;\;\;\;\;\;\;\;\;\;\;\;\;\;\;\;\;\;\;\;\;\;\;\;
+{\left. {{{\pi _{\bf{\theta }}}\left( {{{\boldsymbol{C}}_k}\left| {{{\boldsymbol{H}}_k}} \right.} \right) \cdot {\nabla _{\bf{\theta }}}\left( {Q\left( {{{\boldsymbol{C}}_k},{{\boldsymbol{H}}_k}} \right)} \right)}} \right]},
\label{eq_14}
\end{align}
\end{small}where the gradient of Q function can be written as
\begin{small}
\begin{align}
&{\nabla _\theta }\left( {Q\left( {{{\boldsymbol{H}}_k},{{\boldsymbol{C}}_k}} \right)} \right) = \notag\\
&{\nabla _\theta }\left\{ {\sum\nolimits_{{{\boldsymbol O}_{k + 1}} \in \mathcal{O}} {\Pr \left[ {{{\boldsymbol{H}}_{k + 1}}\left| {{{\boldsymbol{H}}_k},{{\boldsymbol{C}}_k}} \right.} \right] \cdot \left({J_k}\left( {{{\boldsymbol O}_k}} \right) + \lambda  \cdot V\left( {{{\boldsymbol{H}}_{k + 1}}} \right)\right)} } \right\} \notag \\
& = \lambda  \cdot \sum\nolimits_{{{\boldsymbol O}_{k + 1}} \in \mathcal{O}} {\Pr \left[ {{{\boldsymbol{H}}_{k + 1}}\left| {{{\boldsymbol{H}}_k},{{\boldsymbol{C}}_k}} \right.} \right] \cdot {\nabla _\theta }\left( {V\left( {{{\boldsymbol{H}}_{k + 1}}} \right)} \right).} 
\label{eq_15}
\end{align}
\end{small}Furthermore, for arbitrary policy $\pi_{\boldsymbol{\theta}}$, based on the employment of log-derivative trick, we have\begin{small}
\begin{align}
    &{\nabla _{\boldsymbol{\theta}} }\left( {{\pi _{\boldsymbol{\theta}} }\left( {{{\boldsymbol C}_k}\left| {{{\boldsymbol H}_k}} \right.} \right)} \right) = \nonumber \\
    &\;\;\;\;{\pi _{\boldsymbol{\theta}} }\left( {{{\boldsymbol C}_k}\left| {{{\boldsymbol H}_k}} \right.} \right) \cdot \log \left( {{\nabla _{\boldsymbol{\theta}} }\left( {{\pi _{\boldsymbol{\theta}} }\left( {{{\boldsymbol C}_k}\left| {{{\boldsymbol H}_k}} \right.} \right)} \right)} \right). \label{eq_16}
\end{align}
\end{small}Substituting equations (\ref{eq_15}) and (\ref{eq_16}) into gradient (\ref{eq_14}), then \begin{small}${\nabla _{\boldsymbol{\theta}} }V\left( {{{\boldsymbol{H}}_k}} \right)$\end{small}can be rewritten as formula (\ref{eq_17}), where
\begin{figure*}[h]
\begin{small}
\begin{align}
&{{\nabla}_{\boldsymbol{\theta}} }V\left( {{{\boldsymbol{H}}_k}} \right) 
= \sum\nolimits_{{{\boldsymbol{C}}_k} \in \mathcal{C}} {\left[ {{\pi _{\boldsymbol{\theta}} }\left( {{{\boldsymbol{C}}_k}\left| {{{\boldsymbol{H}}_k}} \right.} \right) \cdot \phi \left( {{{\boldsymbol{H}}_k},{{\boldsymbol{C}}_k}} \right)} \right.} + 
\left. {\lambda  \cdot \sum\nolimits_{{{\boldsymbol O}_{k + 1}} \in \mathcal{O}} {{\pi _{\boldsymbol{\theta}} }\left( {{{\boldsymbol{C}}_k}\left| {{{\boldsymbol{H}}_k}} \right.} \right) \cdot \Pr \left[ {{{\boldsymbol{H}}_{k + 1}}\left| {{{\boldsymbol{H}}_k},{{\boldsymbol{C}}_k}} \right.} \right] \cdot {{\nabla}_{\boldsymbol{\theta}} }\left( {V\left( {{{\boldsymbol{H}}_{k + 1}}} \right)} \right)} } \right]
\label{eq_17}
\end{align}
\end{small}
\begin{small}
\begin{align}
{{\nabla}_{\boldsymbol{\theta}} }V\left( {{{\boldsymbol{H}}_k}} \right)
&= \sum\nolimits_{{{\boldsymbol{C}}_k} \in \mathcal{C}} {{\pi _{\boldsymbol{\theta}} }\left( {{{\boldsymbol{C}}_k}\left| {{{\boldsymbol{H}}_k}} \right.} \right) \cdot \phi \left( {{{\boldsymbol{H}}_k},{{\boldsymbol{C}}_k}} \right)} \notag\\
&\;\;\;+ \;\lambda  \cdot \sum\nolimits_{{{\boldsymbol O}_{k + 1}} \in \mathcal{O}} {\Pr \left[ {{{\boldsymbol{H}}_{k + 1}}\left| {{{\boldsymbol{H}}_k};{\pi _{\boldsymbol{\theta}} }} \right.} \right] \cdot \left( {\sum\nolimits_{{{\boldsymbol{C}}_{k + 1}} \in \mathcal{C}} {{\pi _{\boldsymbol{\theta}} }\left( {{{\boldsymbol{C}}_{k + 1}}\left| {{{\boldsymbol{H}}_{k + 1}}} \right.} \right) \cdot \phi \left( {{{\boldsymbol{H}}_{k + 1}},{{\boldsymbol{C}}_{k + 1}}} \right)} } \right)} \notag\\
&\;\;\;+ {\lambda ^2} \cdot \sum\nolimits_{{{\boldsymbol O}_{k + 1}},{{\boldsymbol O}_{k + 2}} \in \mathcal{O}{\rm{,}}{{\boldsymbol{C}}_{k + 1}} \in \mathcal{C}} {\Pr \left[ {{{\boldsymbol{H}}_{k + 2}}\left| {{{\boldsymbol{H}}_k};{\pi _{\boldsymbol{\theta}} }} \right.} \right] \cdot } \left( {\sum\nolimits_{{{\boldsymbol{C}}_{k + 2}} \in \mathcal{C}} {{\pi _{\boldsymbol{\theta}} }\left( {{{\boldsymbol{C}}_{k + 2}}\left| {{{\boldsymbol{H}}_{k + 2}}} \right.} \right) \cdot \phi \left( {{{\boldsymbol{H}}_{k + 2}},{{\boldsymbol{C}}_{k + 2}}} \right)} } \right)
+  \cdots \notag\\
&= \mathbb{E}{_{\tau  > k}}\left[ {\left. {\sum\nolimits_{k' = k}^\infty  {{\lambda ^{k' - k}}{\nabla _{\boldsymbol{\theta}} }\left( {\log \left( {{{\pi}_{\boldsymbol{\theta}} }\left( {{{\boldsymbol{C}}_{k'}}\left| {{{\boldsymbol{H}}_{k'}}} \right.} \right)} \right)} \right) \cdot Q\left( {{{\boldsymbol{H}}_{k'}},{{\boldsymbol{C}}_{k'}}} \right)} } \right|{{\boldsymbol{H}}_k};{{\pi}_{\boldsymbol{\theta}} }} \right] \label{eq_18}
\end{align}
\end{small}
\hrulefill
\end{figure*}
\begin{small}
\[\phi \left( {{{\boldsymbol{H}}_k},{{\boldsymbol{C}}_k}} \right) = {\nabla _{\boldsymbol{\theta}} }\left( {\log \left( {{\pi _{\boldsymbol{\theta}} }\left( {{{\boldsymbol{C}}_k}\left| {{{\boldsymbol{H}}_k}} \right.} \right)} \right)} \right) \cdot Q\left( {{{\boldsymbol{H}}_k},{{\boldsymbol{C}}_k}} \right).\]
\end{small}Equation (\ref{eq_17}) has a nice recursive form and the future state value function \begin{small}$V\left( {{{\boldsymbol{H}}_k'}} \right)$\end{small} (\begin{small}$k'=k+1,k+2,...$\end{small}) can be repeated unrolled by following the same equation.
\par
Then, we keep on unrolling \begin{small}$V\left( {{{\boldsymbol{H}}_k'}} \right)$\end{small} in equation (\ref{eq_17}), we can obtain equation (\ref{eq_18}), where
\begin{small}
\begin{align}
    &\Pr \left[ {{{\boldsymbol{H}}_{k'}}\left| {{{\boldsymbol{H}}_k};{\pi _\theta }} \right.} \right] = \nonumber \\
    &\;\;\;\;\prod\nolimits_{i = 0}^{k' - k} {\pi \left( {\left. {{{\boldsymbol{C}}_{k + i}}} \right|{{\boldsymbol{H}}_{k + i}}} \right)}  \cdot \Pr \left[ {{{\boldsymbol O}_{k + i + 1}}\left| {{{\boldsymbol{H}}_{k + i}},{{\boldsymbol{C}}_{k + i}}} \right.} \right]
\end{align}\end{small}is the probability of transitioning from \begin{small}${\boldsymbol{H}}_k$\end{small} to \begin{small}${\boldsymbol{H}}_k'$\end{small}.
\par
Direct use of equation (\ref{eq_18}) to estimate \begin{small}${{\nabla}_{\boldsymbol{\theta}}}V\left( {{{\boldsymbol{H}}_k}} \right)$\end{small} will induce high variance of gradient estimation. To deal with this issue, researchers propose an idea is to add a ``baseline” that will not affect the expectation but reduce the variance. One such ``baseline” can be derived using following reasoning:

For any policy $\pi_{\boldsymbol{\theta}}$, it is true that \begin{small}$\sum\nolimits_{{{\boldsymbol{C}}_{k'}} \in \mathcal{C}} { {{{\pi}_{\theta}}\left( {{{\boldsymbol{C}}_{k'}}\left| {{{\boldsymbol{H}}_{k'}}} \right.} \right)}} = 1$\end{small}. Then, taking the gradient ${{\nabla}_{\boldsymbol{\theta}}}$ from both sides and utilizing equation (\ref{eq_16}), we can obtain:
\begin{small}
\begin{align}
0 &= \sum\limits_{{{\boldsymbol{C}}_{k'}} \in \mathcal{C}} {{\nabla _{\boldsymbol{\theta}} }\left( {{\pi _{\boldsymbol{\theta}} }\left( {{{\boldsymbol{C}}_{k'}}\left| {{{\boldsymbol{H}}_{k'}}} \right.} \right)} \right)} \notag \\
 & = \sum\limits_{{{\boldsymbol{C}}_{k'}} \in \mathcal{C}} {{\pi _{\boldsymbol{\theta}} }\left( {{{\boldsymbol{C}}_{k'}}\left| {{{\boldsymbol{H}}_{k'}}} \right.} \right) \cdot {\nabla _{\boldsymbol{\theta}} }\left( {\log \left( {{\pi _{\boldsymbol{\theta}} }\left( {{{\boldsymbol{C}}_{k'}}\left| {{{\boldsymbol{H}}_{k'}}} \right.} \right)} \right)} \right)}. \label{eq_19} 
\end{align}
\end{small}Multiplying the expression (\ref{eq_19}) with some value independent of \begin{small}${\boldsymbol C}_{k'}$\end{small}, e.g., \begin{small}$\lambda^{k'-k}V(\boldsymbol{H}_{k'})$\end{small}, we have
\begin{small}
\begin{align}
    \sum\limits_{{{\boldsymbol{C}}_{k'}} \in \mathcal{C}} {{\pi _{\boldsymbol{\theta}} }\left( {{{\boldsymbol{C}}_{k'}}\left| {{{\boldsymbol{H}}_{k'}}} \right.} \right){\nabla _{\boldsymbol{\theta}} }\left( {\log \left( {{\pi _{\boldsymbol{\theta}} }\left( {{{\boldsymbol{C}}_{k'}}\left| {{{\boldsymbol{H}}_{k'}}} \right.} \right)} \right)} \right)} 
    {\lambda ^{k' - k}}V\left( {{{\boldsymbol{H}}_{k'}}} \right)
    = 0. \notag
\end{align}
\end{small}Adding this equation into gradient formula (\ref{eq_18}), \begin{small}${{\nabla}_{\boldsymbol{\theta}}}V\left( {{{\boldsymbol{H}}_k}} \right)$\end{small} can be rewritten as follows,
\begin{footnotesize}
\[\begin{array}{l}
{\nabla _{\boldsymbol{\theta}} }V\left( {{{\boldsymbol{H}}_k}} \right) = \\
 \mathbb{E}{_{\tau  > k}}\left[ {\left. {\sum\limits_{k' = k}^\infty  {{\lambda ^{k' - k}}{\nabla _{\boldsymbol{\theta}} }\left( {\log \left( {{\pi _{\boldsymbol{\theta}} }\left( {{{\boldsymbol{C}}_{k'}}\left| {{{\boldsymbol{H}}_{k'}}} \right.} \right)} \right)} \right) \cdot A\left( {{{\boldsymbol{H}}_{k'}},{{\boldsymbol{C}}_{k'}}} \right)} } \right|{{\boldsymbol{H}}_k};{\pi _{\boldsymbol{\theta}} }} \right],
\end{array}\]
\end{footnotesize}where \begin{small}$A(\boldsymbol{H}_{k'},\boldsymbol{C}_{k'})$\end{small} is the advantage function defined in (\ref{eq_13}a). Then the gradient of the objective function $J(\pi_{\boldsymbol{\theta}})$ is\begin{footnotesize}
\begin{align}
&{\nabla _{\boldsymbol{\theta}} }J\left( {{\pi _{\boldsymbol{\theta}} }} \right)=  \notag\\
&\mathbb{E}{_\tau }\left[ {\left. {\sum\limits_{k' = 1}^\infty  {{\lambda ^{k' - 1}}{\nabla _{\boldsymbol{\theta}} }\left( {\log \left( {{\pi _{\boldsymbol{\theta}} }\left( {{{\boldsymbol{C}}_{k'}}\left| {{{\boldsymbol{H}}_{k'}}} \right.} \right)} \right)} \right) \cdot A\left( {{{\boldsymbol{H}}_{k'}},{{\boldsymbol{C}}_{k'}}} \right)} } \right|{\pi _{\boldsymbol{\theta}}}} \right]. \notag
\end{align}
\end{footnotesize}Then, we obtain \textbf{Theorem 1}.
\end{IEEEproof}

\bibliographystyle{IEEEtran}
\bibliography{IEEEabrv,Ref}
%
%
%
\begin{IEEEbiography}[{{\includegraphics[width=1in,height=1.25in,clip,keepaspectratio]{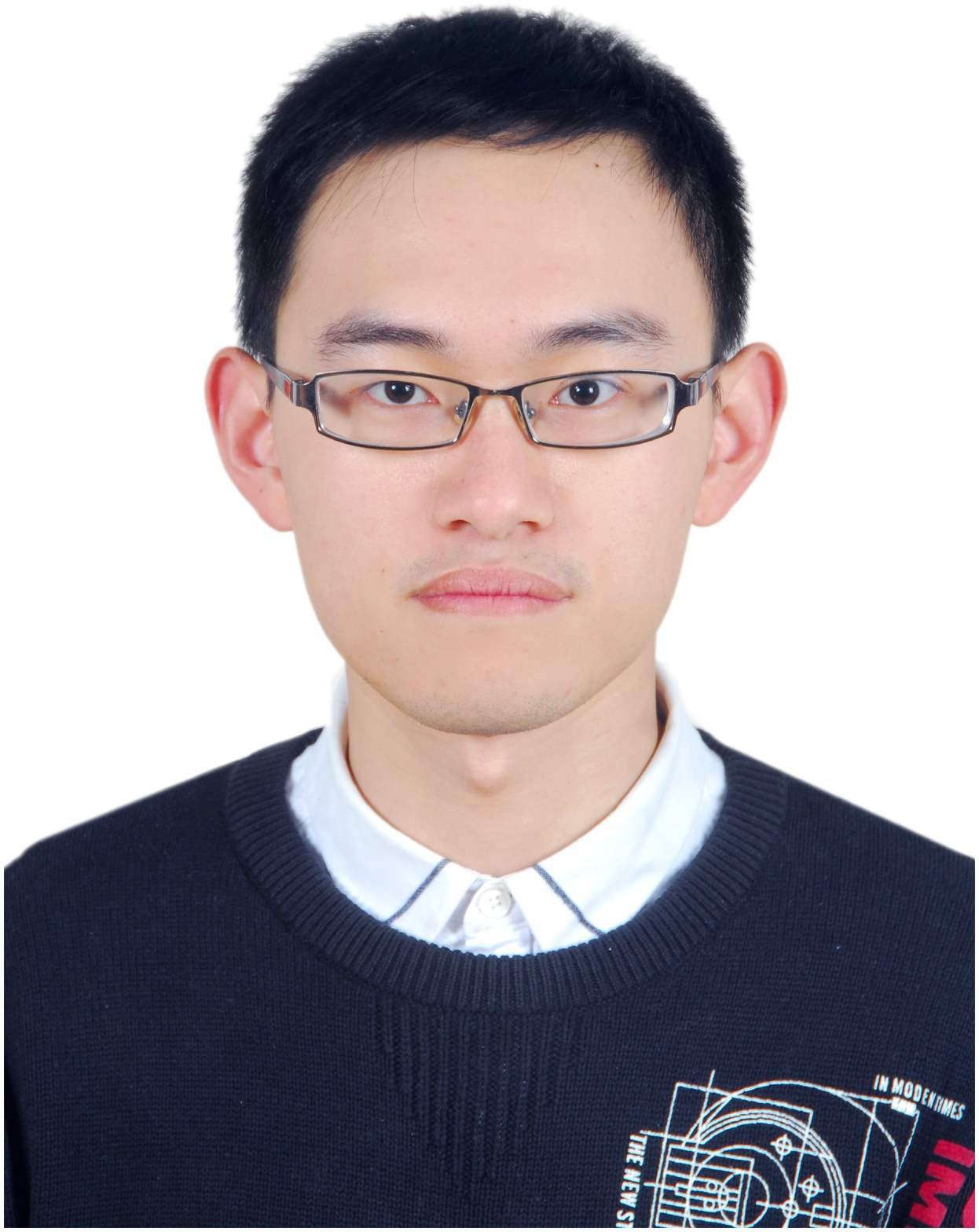}}}]
{Jie Mei} (S'18-M'19) received the B.S. degree from the Nanjing University of Posts and Telecommunications (NJUPT), China, in 2013, and the Ph.D. degree in information and communication engineering from the Beijing University of Posts and Telecommunications (BUPT) in June 2019. He is currently a Post-Doctoral Associate with the Electrical and Computer Engineering, Western University, Canada. His research interests include intelligent communications, multi-dimensional intelligent multiple access, and Vehicle-to-Everything (V2X) communication. He was a TPC member of IEEE Globecom 2020. 
\end{IEEEbiography}

\begin{IEEEbiography}[{{\includegraphics[width=1in,height=1.25in,clip,keepaspectratio]{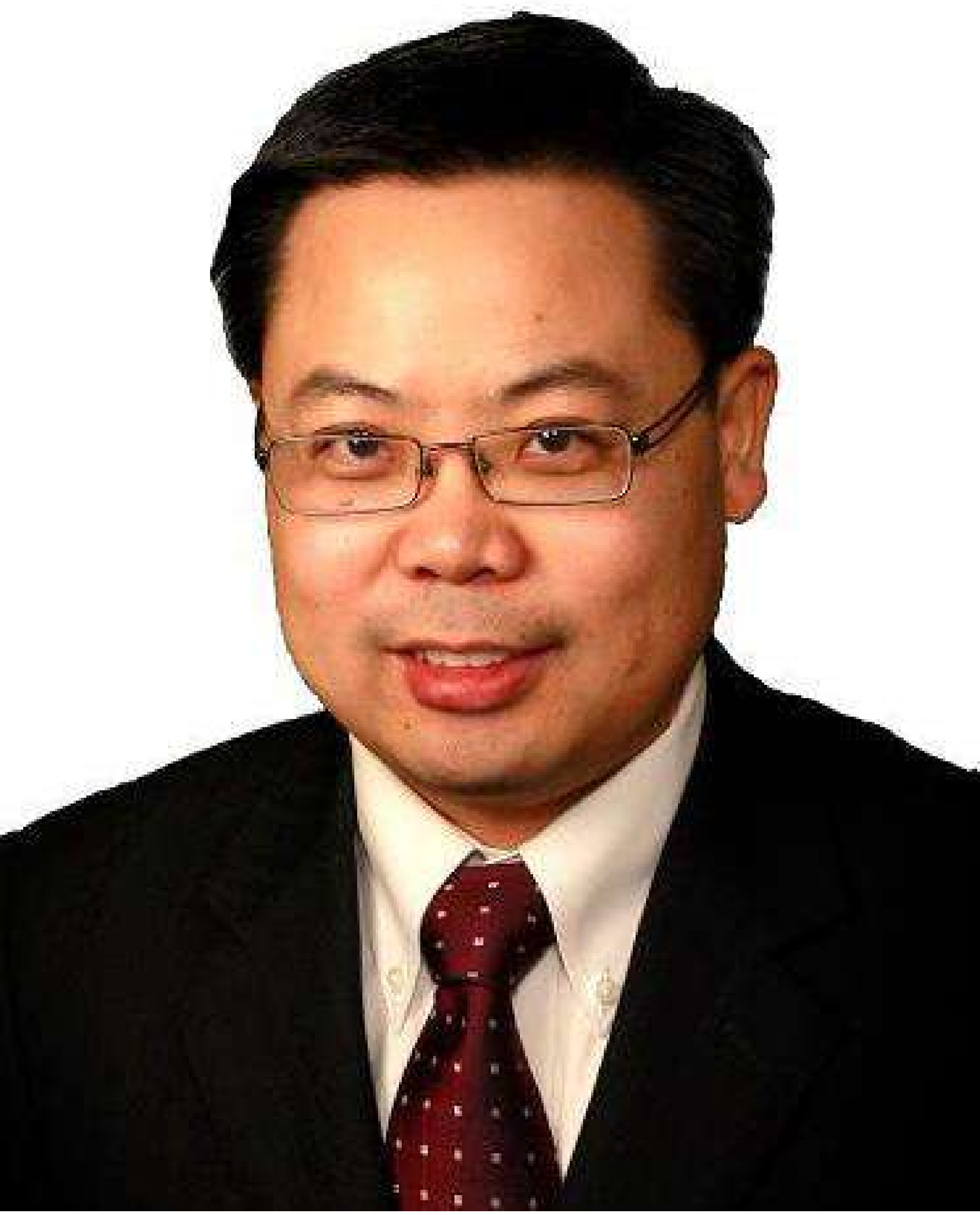}}}] 
{Dr. Xianbin Wang} (S'98-M'99-SM'06-F'17) is a Professor and Tier-1 Canada Research Chair at Western University, Canada. He received his Ph.D. degree in electrical and computer engineering from the National University of Singapore in 2001.

Prior to joining Western, he was with Communications Research Centre Canada (CRC) as a Research Scientist/Senior Research Scientist between July 2002 and Dec. 2007. From Jan. 2001 to July 2002, he was a system designer at STMicroelectronics.  His current research interests include 5G/6G technologies, Internet-of-Things, communications security, machine learning and intelligent communications. Dr. Wang has over 450 highly cited journal and conference papers, in addition to 30 granted and pending patents and several standard contributions.

Dr. Wang is a Fellow of Canadian Academy of Engineering, a Fellow of Engineering Institute of Canada, a Fellow of IEEE and an IEEE Distinguished Lecturer. He has received many awards and recognitions, including Canada Research Chair, CRC President’s Excellence Award, Canadian Federal Government Public Service Award, Ontario Early Researcher Award and six IEEE Best Paper Awards. He currently serves/has served as an Editor-in-Chief, Associate Editor-in-Chief, Editor/Associate Editor for over 10 journals. He was involved in many IEEE conferences including GLOBECOM, ICC, VTC, PIMRC, WCNC, CCECE and CWIT, in different roles such as general chair, symposium chair, tutorial instructor, track chair, session chair, TPC co-chair and keynote speaker. He has been nominated as an IEEE Distinguished Lecturer several times during the last ten years. Dr. Wang is currently serving as the Chair of IEEE London Section and the Chair of ComSoc Signal Processing and Computing for Communications (SPCC) Technical Committee.
\end{IEEEbiography}

\begin{IEEEbiography}[{{\includegraphics[width=1in,height=1.25in,clip,keepaspectratio]{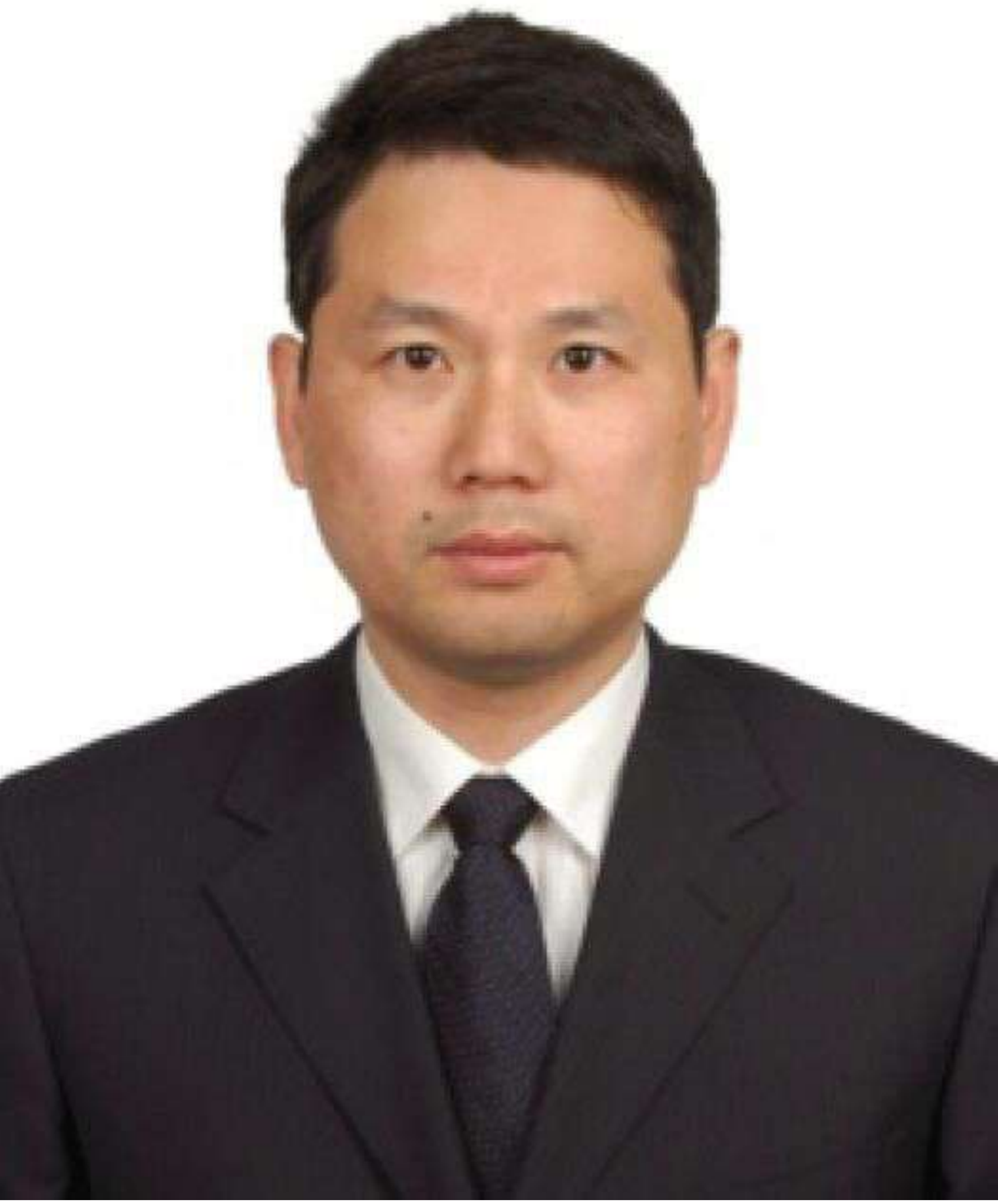}}}]
{Kan Zheng} (S'02-M'06-SM'09) received the B.S., M.S., and Ph.D. degrees from Beijing University of Posts and Telecommunications (BUPT), China, in 1996, 2000, and 2005, respectively, where he is currently a Professor. He has rich experiences on the research and standardization of the new emerging technologies. He is the author of more than 200 journal articles and conference papers in the field of wireless networks, IoT, vehicular communication, and so on. He holds editorial board positions for several journals and has organized several special issues, including IEEE COMMUNICATIONS SURVEYS \& TUTORIALS, IEEE Communication Magazine, and IEEE SYSTEM JOURNAL.
\end{IEEEbiography}
\end{document}